\documentclass[twocolumn,twocolappendix]{aastex63}
\usepackage{float}

\newcommand{\gppr}{\stackrel{>}{\scriptstyle \sim}}

\accepted{2020 April 30}

\shorttitle{Gap-type Particle Acceleration in Rotating BH Magnetospheres}
\shortauthors{Katsoulakos \& Rieger}

\begin{document}

\title{Gap-type Particle Acceleration in the Magnetospheres of Rotating Supermassive 
Black Holes}

\correspondingauthor{Grigorios Katsoulakos}
\email{gkats@mpi-hd.mpg.de}

\author[0000-0002-8155-5969]{Grigorios Katsoulakos}
\affiliation{Max-Planck-Institut f\"{u}r Kernphysik, P.O. Box 103980, D-69029 Heidelberg, 
Germany}
\affiliation{International Max Planck Research School for Astronomy and Cosmic Physics, 
University of Heidelberg (IMPRS-HD), Germany}
\affiliation{ZAH, Institut f\"{u}r Theoretische Astrophysik, Universit\"{a}t Heidelberg, 
Philosophenweg 12, D-69120 Heidelberg, Germany}

\author[0000-0003-1334-2993]{Frank M. Rieger}
\affiliation{ZAH, Institut f\"{u}r Theoretische Astrophysik, Universit\"{a}t Heidelberg,
Philosophenweg 12, D-69120 Heidelberg, Germany}
\affiliation{Max-Planck-Institut f\"{u}r Kernphysik, P.O. Box 103980, D-69029 Heidelberg, 
Germany} 

\begin{abstract}
The detection of rapidly variable gamma-ray emission in active galactic nuclei (AGN)
has generated renewed interest in magnetospheric particle acceleration and emission 
scenarios. In order to explore its potential, we study the possibility of steady 
gap acceleration around the null surface of a rotating black hole magnetosphere. 
We employ a simplified (1D) description along with the general relativistic 
expression of Gauss's law, and we assume that the gap is embedded in the radiation 
field of a radiatively inefficient accretion flow. The model is used to derive 
expressions for the radial distribution of the parallel electric field component, 
the electron and positron charge density, the particle Lorentz factor, and the number 
density of $\gamma$-ray photons. We integrate the set of equations numerically, 
imposing suitable boundary conditions. The results show that the existence of a
steady gap solution for a relative high value of the global current is in principle
possible if charge injection of both species is allowed at the boundaries. We present 
gap solutions for different choices of the global current and the accretion rate. 
When put in context, our results suggest that the variable very high energy 
$\gamma$-ray emission in M87 could be compatible with a magnetospheric origin.
\end{abstract}


\section{Introduction}\label{sec:01}

The nonthermal processes occurring in the vicinity of supermassive black holes 
(BHs) have attracted considerable attention in recent times \citep[e.g.,][]{hir16,
hir17,lev17,hir18,for18,lev18,kats18,che18,pet19}. The formation of strong 
electromagnetic fields in charge-deficient regions (aka gaps) around rotating BHs 
is thought to facilitate efficient particle acceleration to very high energies (VHEs), 
in the case of hadrons possibly even up to ultrahigh ($\geq 10^{18}$ eV) energies 
\citep[see][for a review]{rie19}. This process is naturally accompanied by 
gamma-ray production via curvature emission and inverse Compton (IC) upscattering 
of ambient (accretion disk) soft photons. Efficient annihilation of gamma-ray 
photons could trigger an electromagnetic cascade, providing a plasma source for 
continuous jet formation \citep{lev11}. Given suitable conditions, the close 
BH environment could enable significant power extraction and account for rapid 
gamma-ray variability on horizon crossing times $r_g/c =1.4~(M_{BH}/10^9\,
M_{\odot})$ hr and shorter \citep{ale14}. It seems possible that the variable 
VHE emission from radio galaxies, and in particular from M87, reveals signs of 
such processes \citep[see][for a recent review]{rie18}. Given its proximity 
(distance $d\simeq 17$ Mpc)\citep{can18}, undeluminosity ($L_{\rm bol} 
\leq 10^{-6} L_{\rm Edd}$) and high BH mass ($M_{\rm BH}=6.5\times 10^9M_{\odot}$)
\citep{aki19a,aki19c}, M87 in fact provides a unique laboratory in this regard
\citep[e.g.,][]{ner07,lev11,pti16,kats18,ait19}.

Gap-type particle acceleration can occur if the available charge density falls 
below a critical value ($\rho_{\rm GJ}$) needed to screen off the (parallel) electric 
field. A generic feature in this context is the occurrence of a specific region 
(referred to as the null surface) in the immediate vicinity of a rotating BH
across which the critical density changes sign and gaps may form \citep{bes92,
hir98}. It has been suggested early on that the ensuing electromagnetic cascades 
could facilitate the charge supply needed to support a force-free jet magnetosphere 
\citep{bla77,mac82}. To understand the dynamics, the resultant electric field
and acceleration, as well as the pair (charge) and photon distributions in the
gap, need to be self-consistently described. In the present paper this is done 
by investigating a simplified (1D) steady gap model following previous approaches
\citep[e.g.,][]{hir98,hir99}. In the current study, two major modifications have 
been implemented in the model. Firstly, we explore numerical solutions of the gap 
structure taking into account the general relativistic expression of Gauss's 
law and applying the relativistic formula of the Goldreich-Julian charge density, 
$\rho_{\rm GJ}$. Secondly, targeting low-luminosity AGNs \citep[e.g.,][]{ho09,xu10,nem14}, 
we assume that the BH is embedded within the radiation field of an optically-thin 
advection-dominated accretion flow (ADAF), so that the ambient soft photon field 
(its strength and relevant energy range) can change significantly with accretion 
rate. We consider that such a gap model provides a useful tool to get physical 
insight into possible characteristics of magnetospheric gamma-ray emission in AGNs. 
For a full relativistic treatment of steady gap accelerators, the reader is 
referred to recently published studies \citep{hir16,hir17,lev17}. As we show below, 
however, the implementation of the relativistic Goldreich-Julian charge density 
$\rho_{\rm GJ}$ seems sufficient to capture the relevant information. 

One expects a steady gap approach to be an idealization, as gap formation could 
well be intermittent \citep[e.g.,][]{lev17}. Recent PIC simulations by \citet{lev18},
\citet{che18} and \citet{che19}, however, do not yet agree on the overall characteristics 
and apply simplified descriptions for the ambient soft photon field. The approach chosen 
here seems beneficial in that it allows us to get some first insights into possible 
dependencies of the gap structure on different and more complex ambient soft photon 
fields. This remains relevant even if the ultimate regulation mechanisms for 
intermittent gaps were to be different.

The paper is structured as follows: Sec.~2 introduces the general framework, 
while Sec.~3 describes the system of equations governing the gap accelerator. 
Suitable normalization and boundary conditions are discussed in Sections~4 and 5. 
Constraints on the existence of steady gap solutions are explored in Sec.~6.
The numerical method and selected solutions are then described in Sec.~7 and 
Sec.~8. An application to M87 is finally discussed in Sec.~9.

\section{Model framework}\label{sec:02}
\subsection{Black Hole Vicinity}
We consider a rotating BH of mass $M=M_9 \times 10^9 M_{\odot}$ and angular 
momentum $J=GM^2/c$ onto which gas accretion occurs. The BH is fed by the 
accretion flow at a rate $\dot{M}=\dot{m}\,\dot{M}_{\rm Edd}$ expressed in Eddington 
units (assuming a standard conversion efficiency), where $\dot{M}_{\rm Edd} \approx 
1.4\times 10^{27} M_{9}$ $g\, s^{-1}$. Provided that the disk supports a large-scale 
magnetic field \citep[for a review]{haw15}, this is expected to follow the inward motion 
of gas and to accumulate in the immediate vicinity of the BH. The characteristic 
magnetic field strength close to the horizon is of order \citep{kats18}
\begin{equation}
B_H \simeq 10^5\,\dot{m}^{1/2}\,M_{9}^{-1/2}\quad  \mathrm{G}.    
\end{equation}

\begin{figure}[htb]
\hspace{-0.4cm}
\includegraphics[width=0.49\textwidth,height=7.7cm]{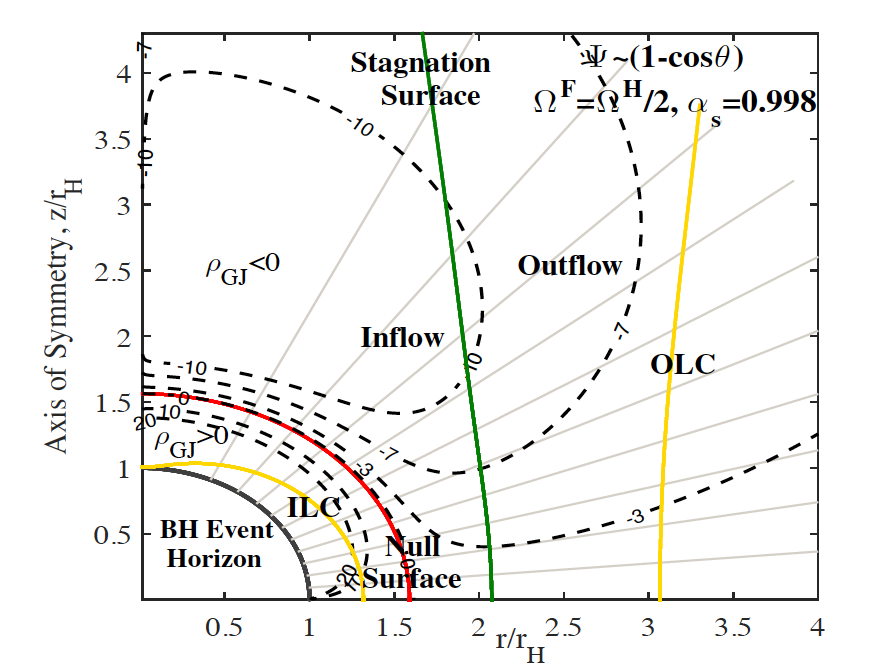}
\caption{Poloidal representation of a steady BH magnetosphere for a split 
monopole magnetic field configuration (gray lines). Potential sites for efficient 
particle acceleration, namely, the null surface (red line) and the stagnation surface 
(green line), are located between the inner and outer light cylinder (ILC  and OLC; 
yellow lines). Contour levels of the charge density (dashed lines) are shown with
dashed lines. We note that the null surface and the stagnation surface are related 
to the sign change of the charge density and the neutral matter separation, 
respectively.}
    \label{topology}
\end{figure}

We further consider the existence of a plasma source (provided by, e.g. 
$\gamma\gamma$-annihilation of disk photons or some electromagnetic cascade) capable 
of filling the BH magnetosphere with a sufficient amount of charged particles. 
In addition, we assume that the black hole rotation $\Omega^{H}$, the magnetic field 
$\textbf{B}_{H}$ and the amount of charges $\rho_{e}$ are such that they can ensure 
degeneracy (i.e., $\textbf{E}\cdot\textbf{B}=0$) and force-freeness (i.e., $\rho_{e}
\textbf{E}+(\textbf{j}/c)\times\textbf{B}=0$) almost everywhere in the magnetosphere. 
It is known that a force-free magnetosphere leads to efficient extraction of the 
rotational energy of the BH, facilitating jet or outflow formation \citep{bla77}. 
The associated Blandford-Znajek-type jet power is 
\begin{eqnarray}\label{L_BZ}
    L_{BZ} =\Omega^{F}\left(\Omega^{H}-\Omega^{F}\right)B_{\perp}^2\frac{r_{H}^4}{c} 
    \qquad \qquad \qquad \\ \nonumber 
    \approx 2\times 10^{48}\,\dot{m}\, M_{9} \quad  \mathrm{erg\, s^{-1}},
\end{eqnarray}
where $\Omega^{F}=\Omega^{H}/2$ is the angular velocity of the magnetic field lines 
and $B_{\perp}\approx B_{H}$ is the magnetic field strength that threads the horizon. 
Even under these circumstances, however, the emergence of electric field components 
$E_{||}$ across the null surface parallel to the magnetic field (i.e., gap acceleration) 
is possible, since continuous charge replenishment is required (see, e.g.,  
Fig.~\ref{topology}). 

Hence, the gap accelerator is confined to a region that contains large-scale 
electromagnetic fields, charged particles (i.e., either in surplus in the force-free 
domain $\rho_{e}\geq \rho_{\rm GJ}$, or in deficit within the accelerating zone 
$\rho_{e}<\rho_{\rm GJ}$) and ambient soft photons associated with emission from 
the disk. In what follows, we adopt a split monopole topology
\begin{equation}
    \Psi= 4\pi r_{H}^2B_{H}(1-\cos\theta)\,,
\end{equation}
where $\theta$ gives the angle with respect to the polar axis, and $r_H$ is the event 
horizon radius. 

We utilize the "$3+1$"-formalism in the following, according to which the 4D 
spacetime around a rotating BH splits into 3D space, i.e., absolute 
space, and 1D time, i.e, the global time $t$, \citep[for details see,][]{tho86}. 
The absolute space is described, using a Boyer-Lindquist spatial coordinate system 
(r,$\theta$,$\phi$), by the metric \citep{tho86}
\begin{equation}
ds^2=\gamma_{rr}dr^2+\gamma_{\theta\theta}d\theta^2+\gamma_{\phi\phi}d\phi^2, \label{eq2.01}
\end{equation}
where $\gamma_{ij}$ are the matrix elements of the space-metric tensor given by
\begin{eqnarray}
\gamma_{ij} & = & 
\left(\begin{array}{ccc} 
 \frac{\rho^2}{\Delta} & 0 & 0\\ 
  0 & \rho^{2} & 0\\
  0 & 0 & \widetilde{\omega}^2 
\end{array}\right),            \label{eq3.02}               
\end{eqnarray} 
with $\rho$, $\Delta$ and $\widetilde{\omega}$ given by the expressions
\begin{equation}
\rho^{2}=r^2+\alpha_{s}^2\cos^2{\theta},\label{eq3.03}
\end{equation}
\begin{equation}
 \Delta=r^2-2r_{g}r+\alpha_{s}^2,\label{eq3.04}
\end{equation}
\begin{equation}
\widetilde{\omega}=\frac{\Sigma}{\rho}\sin\theta, \label{eq3.05}
\end{equation}
and the function $\Sigma$ given by
\begin{equation}
 \Sigma^{2}=(r^2+\alpha_{s}^2)^{2}-\alpha_{s}^2\Delta\sin^2{\theta}.\label{eq3.06}
\end{equation}
In addition, we have defined the gravitational radius, $r_{g}=GM/c^2$, and the 
spin parameter of the BH, $\alpha_{s}=J/Mc$. The inverse matrix of the 
metric tensor is given by $\gamma^{ij}=\left(\Delta/\rho^2,1/\rho^2,
1/\tilde{\omega}^2\right)$.

In the "$3+1$" formalism all the laws and physical quantities are measured by 
fiducial observers (FIDOs), for Kerr BHs also often referred to as "zero 
angular momentum observers" (ZAMOs), carrying their own clocks and located in 
each point of absolute space. 
Given that the BH rotates and drags all the physical objects near it, FIDOs 
must also have a radius-dependent, finite, angular velocity relative to absolute space
\begin{equation}
 \left.\frac{d\phi}{dt}\right|_{\rm FIDO}=-\beta^{\phi}=\omega.\label{eq3.07}
\end{equation}
Furthermore, the gravity of the BH causes a gravitational redshift to their 
clocks. Their lapse of proper time $d\tau$ is related to the lapse of the global 
time $dt$ via the function
\begin{equation}
\left.\frac{d\tau}{dt}\right|_{\rm FIDO}=\alpha_{l}. \label{eq3.08}
\end{equation}
Evidently, in "$3+1$" splitting, general relativistic effects become apparent via 
the so-called Lapse function and Lense-Thirring angular velocity
\begin{equation}
 \alpha_{l}=\frac{\rho\sqrt{\Delta}}{\Sigma}, \qquad  \omega
 =\frac{2c\alpha_{s}r_{g}r}{\Sigma^2}. \label{eq3.09}
\end{equation}
Finally, imposing $\Delta=0$, we find the event horizon radius
\begin{equation}
 r_{H}=r_{g}+\sqrt{r_{g}^2-\alpha_{s}^2},  \label{eq3.10}
\end{equation}
and the event horizon angular velocity (thereafter, the angular velocity of the 
BH)
\begin{equation}
 \Omega^{H}=\frac{c\alpha_{s}}{2r_{g}r_{H}}.\label{eq3.11}
\end{equation}
The spin $a_s$ is in the following expressed in terms of a dimensionless spin 
parameter $a_s^* = a_s/r_g$.

\begin{figure}[ht!]
\includegraphics[width=0.49\textwidth,height=7.5cm]{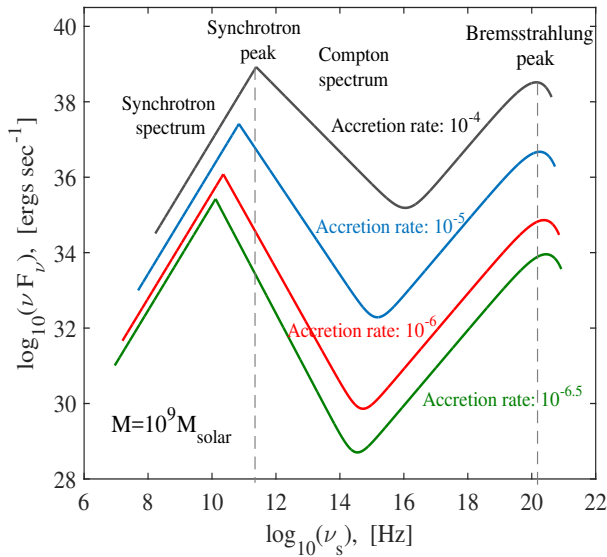}
\caption{Characteristic ADAF spectra for accretion rates $\dot{m}=10^{-4}$ (black line), 
$\dot{m}=10^{-5}$ (blue line), $\dot{m}=10^{-6}$ (red line), and $\dot{m}=10^{-6.5}$ 
(green line), respectively. A BH of mass $M_{BH}=10^{9}M_{\odot}$ has been 
employed.}
\label{ADAF}
\end{figure}

\subsection{Ambient Soft Photon Field}
We adopt a simplified, optically thin ADAF spectrum \citep[e.g.,][]{nar95a,nar95b} as 
the characteristic soft photon field in our model. This description provides a convenient 
approximation to underluminous AGNs, of which M87 is a prototype. As we have shown 
elsewhere, a radiatively inefficient accretion flow is in fact a prerequisite for the 
observability of magnetospheric VHE emission \citep{kats18}.

Typically, the radio to hard X-€"ray emission in an ADAF is produced by semi-relativistic, 
thermal electrons via synchrotron, IC, and bremsstrahlung processes. 
The synchrotron flux is proportional to \citep{mah97}
\begin{equation}
F_{\nu}^{\rm syn}\propto M_{9}^{6/5}
\dot{m}^{4/5}\,T_{e}^{21/5}\,\nu^{2/5}, 
\label{eq2.62}
\end{equation}
where $T_{e}$ is the temperature of the thermal electrons and $\nu$ is the frequency 
of the emission. As the magnetic field strength depends on mass accretion, the emission 
and the peak frequency vary with accretion rate (see Fig.~\ref{ADAF}). In addition, inverse 
Compton upscattering of the synchrotron photons by the hot electrons 
in the disk produces emission extending up to $h\nu \sim kT_{e}$. The Compton flux above 
the synchrotron peak then approximately follows a power law \citep{mah97}
\begin{equation}
F_{\nu}^{com}=F_{p}^{syn}\left(\frac{\nu_{f}}{\nu_{p}^{syn}}\right)^{-\Gamma}, \label{eq1.63}
\end{equation} 
where $F_{p}^{\rm syn}$ denotes the emission at the peak frequency $\nu_{p}^{\rm syn}$. 
In Fig.~(\ref{ADAF}) four ADAF spectra are shown for different values of the accretion rate. 
The spectra are calculated following the approach of \citet{mah97}. We use these spectra 
and, more specifically, the synchrotron and Compton components,
to determine the soft photon number density per unit energy (i.e, $dN_{s}/dE_{s}$) needed 
as input in the steady gap model below.

\section{The governing equations} \label{sec:03}
It is worth recapturing, at this point, the physics of the gap mechanism \citep[e.g.,][]{lev11}. 
Seed leptons $e^{\pm}$ injected into the gap are quasi-instantaneously accelerated along the 
parallel electric field component. Their energy saturates owing to inverse Compton and curvature 
emission. The resultant $\gamma$-ray photons undergo $\gamma\gamma$-annihilation with 
soft photons of the accretion disk, providing extra leptons to the gap. These secondary 
leptons are then also subjected to acceleration and $\gamma$-ray emission. Correspondingly, 
the secondary $\gamma$-ray photons produce the next generation of pairs, which, in turn, 
radiate the next generation of photons, and so on. In such a way, an electromagnetic 
cascade is triggered and ends only when the charge density $\rho_{e}$ reaches the 
Goldreich-Julian one, $\rho_{\rm GJ}$ \citep{gol69}.  
 
Below, we introduce the system of equations that determines the structure of a 1D
steady gap accelerator. This includes expressions for the radial distribution of the parallel 
electric field component, the Lorentz factor of the particles, the charge density of electrons 
and positrons, as well as the number density of $\gamma$-ray photons.

\subsection{The parallel electric field} \label{subsec:03.01}
The BH rotation and the nearby presence of a magnetic field result in the appearance of 
a large-scale electric field. Given that charged particles move along magnetic 
field lines, the electric field component relevant for acceleration is the parallel one. 

Our reference point is Gauss's law, which relates the electric field to the charge density. In 
the "$3+1$" formalism \citep{tho82}, the form of Gauss's law is similar to 
the classical one,
\begin{equation}
\nabla\cdot \textbf{E}=4\pi\rho_{e}, \label{eq3.12}
\end{equation}
where $\textbf{E}$ and $\rho_{e}$ are the electric field and the charge density, respectively, 
as measured (in units of proper time $\tau$) by ZAMOs. Assume now that one wishes to transform 
the electric field from the coordinate system of ZAMO to a frame comoving with the field lines. 
It is very instructive to think that ZAMO observers play a role equivalent to those of laboratory 
frames in special relativity. Hence, it is sufficient to apply a Lorentz transformation, so that 
the comoving electric field (in units of global time $t$) becomes
\begin{eqnarray}
\mathcal{E}_{||}=\gamma^{F}\left(\alpha_{l}\,\textbf{E}+\frac{\textbf{v}^{F}}{c}\times\alpha_{l}\,
\textbf{B}\right) \approx \nonumber  \\
  \approx \alpha_{l}\,\textbf{E}+\frac{(\Omega^{F}-\omega)}{2\pi c}\, \nabla \Psi,\label{eq3.13}
 \end{eqnarray}
where $\alpha_{l}\,\textbf{E}$ and $\alpha_{l}\,\textbf{B}$ are the electric and 
magnetic field as measured (in units of global time $t$) by a ZAMO frame, $\gamma^{F}=1/
\sqrt{1-(\textbf{v}^{F}/c)^2}$ is the Lorentz factor (here $\gamma^{F}\approx 1$) and
$\textbf{v}^{F}=(1/\alpha_{l})(\Omega^{F}-\omega)\,\widetilde{\omega}\,\textbf{e}_{\hat{\phi}}$ 
is the field line velocity (measured by ZAMO in units of proper time $\tau$), with  $\Omega^{F}$
the angular velocity of the field line, $\widetilde{\omega}$ the cylindrical radius and
$\textbf{e}_{\hat{\phi}}$ the unit vector in the $\phi$-direction. The second term in equation
(\ref{eq3.13}) describes the electric field of a degenerate, force-free, and stationary BH
magnetosphere \citep{tho86}
\begin{equation}
\textbf{E}^{ff}=-\frac{(\Omega^{F}-\omega)}{2\pi\,\alpha_{l}\, c}\, \nabla\Psi. \label{eq3.14} 
\end{equation} If $\mathcal{E}_{||}=0$ everywhere in space, then the field is given by equation 
(\ref{eq3.14}). In that case, the electric field is purely perpendicular to the field lines and 
particle acceleration does not occur. On the other hand, for $\mathcal{E}_{||}\neq 0$ somewhere 
in space, charged particles injected into such regions will experience "one-shot" acceleration.

Substituting equation (\ref{eq3.13}) into Gauss's law (\ref{eq3.12}) yields
\begin{equation}
 \nabla\cdot\left(\frac{\mathcal{E}_{||}}{\alpha_{l}}\right)
 +\nabla\cdot\left[-\frac{(\Omega^{F}-\omega)}{2\pi\,\alpha_{l}\, c}\, 
 \nabla \Psi\right]=4\pi\rho_{e}. \label{eq3.15} 
\end{equation} 
The generalized, critical density $\rho_{GJ}$ is given by 
\begin{equation}
 \rho_{GJ}=\frac{1}{4\pi}\nabla\cdot\textbf{E}^{ff}=
 \frac{1}{4\pi}\nabla\cdot\left[-\frac{(\Omega^{F}
 \omega)}{2\pi\,\alpha_{l}\,c}\,\nabla \Psi\right]. 
 \label{eq3.16} 
\end{equation}
In an environment rich of plasma (i.e., $\rho_{e}>\rho_{GJ}$), the ability of charges to 
move freely along magnetic lines will also ensure degeneracy (i.e., $\textbf{E}\cdot
\textbf{B}=0$). In an environment poor of plasma (i.e., $\rho_{e}<\rho_{GJ}$), on the 
other hand, the amount of charges is not sufficient to guarantee full screening of the 
field. Therefore, a parallel electric field component can emerge in charge-sparse regions, 
aka gaps. Substituting this in equation (\ref{eq3.15}) one obtains
\begin{equation}
\nabla\cdot\left(\frac{\mathcal{E}_{||}}{\alpha_{l}}\right)=4\pi(\rho_{e}-\rho_{GJ}), 
\label{eq3.17} 
\end{equation}
where $\nabla\cdot$ now indicates the divergence of a vector in curved space.
Assuming that the BH magnetosphere is axisymmetric (i.e., $\partial/
\partial\phi=0$) and ignoring polar variations (i.e., $\partial/\partial\theta=0$) 
in Gauss's law, equation (\ref{eq3.17}) becomes
\begin{equation}
\frac{1}{\sqrt{|\gamma|}}\frac{\partial}{\partial r}
\left(\sqrt{|\gamma|}\,\frac{\mathcal{E}_{||}^{r}}{\alpha_{l}}\right)
=4\pi(\rho_{e}-\rho_{GJ}),  \label{eq3.18} 
\end{equation}
where $|\gamma|=\rho^4\,\widetilde{\omega}^2/\Delta$ is the determinant of the metric 
$\gamma_{ij}$ in absolute (3D) space. It is worth emphasizing that $\mathcal{E}_{||}^{r}$ 
is the contravariant component of the corotating electric field and not the physical one, that is, 
$\mathcal{E}_{||}^{\hat{r}}$. If one wishes to express the physical component in terms of the 
contravariant one, one has
\begin{equation}
 \mathcal{E}_{||}^{\hat{r}}=\sqrt{\gamma_{rr}}\,\mathcal{E}_{||}^{r}, \label{eq3.19}
\end{equation}
where $\gamma_{rr}$ is the metric element of absolute space. Equation (\ref{eq3.18}) can be 
rearranged as
\begin{equation}
\frac{d}{dr}\left(\frac{\mathcal{E}_{||}^{r}}{\alpha_{l}}\right)
=4\pi(\rho_{e}-\rho_{GJ})-\frac{1}{\sqrt{|\gamma|}}
\frac{d\sqrt{|\gamma|}}{dr}\left(\frac{\mathcal{E}_{||}^{r}}{\alpha_{l}}\right),            
\label{eq3.20} 
\end{equation} where the term $\mathcal{A}:=(1/\sqrt{|\gamma|})(d\sqrt{|\gamma|}/dr)$ can 
be evaluated by applying the derivative over $r$. The remaining element of equation (\ref{eq3.20}), 
that has to be calculated is the Goldreich-Julian charge density $\rho_{\rm GJ}$, equation (\ref{eq3.16}), 
which involves the computation of a 3D Laplacian, i.e. $\nabla_{i}\nabla^{i}$. For the
noted metric, one finds
\begin{eqnarray}
 \nabla^{k}\Psi=\gamma^{rr}\frac{\partial \Psi}{\partial r}\,\textbf{e}_{r}
 +\gamma^{\theta\theta}\frac{\partial \Psi}{\partial\theta}\,\textbf{e}_{\theta}
 = \quad \quad \quad \quad \nonumber \\ 
 =\frac{\Delta}{\rho^2}\frac{\partial \Psi}{\partial r}\,\textbf{e}_{r}
 +\frac{1}{\rho^2}\frac{\partial \Psi}{\partial\theta}\,\textbf{e}_{\theta}. \label{eq3.25} 
 \end{eqnarray}
Accordingly, we have
\begin{equation}
  \nabla^{k}\Psi=\nabla^{\theta}\Psi=4\pi r_{H}^2 B_{H} \frac{\sin\theta}{\rho^2}\,
  \textbf{e}_{\theta},  
  \label{eq3.26} 
\end{equation}
where the resultant vector has no radial dependence, since a split monopole has been assumed. 
Substituting equation (\ref{eq3.26}) in equation (\ref{eq3.16}), the critical charge density 
becomes
\begin{equation}
 \rho_{GJ}=-\frac{B_{H}\,r_{H}^2}{2\pi c} \nabla_{\theta}
 \left[\frac{(\Omega^{F}-\omega)}{\alpha_{l}}\,\frac{\sin\theta}{\rho^2}\,
 \textbf{e}_{\theta}\right],                   
 \label{eq3.27} 
 \end{equation}
where $\nabla_{\theta}=(1/\sqrt{|\gamma|})(\partial/\partial \theta)(\sqrt{|\gamma|}\,\, )$. 
Finally, after some manipulation, the following relation is obtained:
\begin{eqnarray}
 \rho_{GJ}=-\frac{(\Omega^{F}-\omega)\,B_{H}\,
 \cos\theta}{2 \pi\, c\, \alpha_{l}}\left[\frac{2\,r_{H}^2}{\rho^2}
   -\right. \quad \quad \quad \quad \quad \nonumber  \\  
  -2\,\alpha_{s}^2\,\frac{\Delta\, r_{H}^2}{\rho^2\,
  \Sigma^2}\,\sin^2\theta+2\alpha_{s}^2\,\frac{r_{H}^2}{\rho^4}\,\sin^2\theta
  -  \quad \quad \quad \nonumber  \\ 
 \left.-\frac{4c\,\alpha_{s}^3\, r_g\,
 r_{H}^2}{(\Omega^{F}-\omega)}\frac{r\,\Delta}{\rho^2\,\Sigma^4}\,\sin^2\theta\right].
 \label{eq3.28}
\end{eqnarray}
The dominant term, which determines the distribution of the density along the 
$r$-direction, is the first one inside the brackets. Note that for $\alpha_{s}
\approx 0$, equation(\ref{eq3.28}) reduces to the expression calculated by Beskin 
in the limit of slow hole rotation \citep{bes10}.  

\subsection{The equation of motion} \label{subsec:03.02}
Electrons and positrons moving into the gap will experience an efficient "one-shot" 
acceleration. The particle Lorentz factor will quickly reach its maximum value, where
acceleration is balanced by energy losses. Without loss of generality we can assume that
$\mathcal{E}_{||}^{r}<0$, namely, the electric field points toward the BH. This
configuration is normally realised, if the axis of black hole rotation is aligned with 
the magnetic one (see, e.g., equation \ref{eq3.14}). 
As a consequence of this field direction, electrons move outward while positrons move inward, 
creating a charge species asymmetry across the gap boundaries. Moreover, the change of charge 
sign compensates the change of velocity sign, resulting in a common equation of motion for 
both species. Hence, the equation that describes the motion of leptons (both electrons and
positrons) within the gap is given by
\begin{equation}
 m_{e}c^2\frac{d\Gamma_{e}}{dr}=-e\mathcal{E}_{||}^{r}-\frac{P_{IC}}{c}-\frac{P_{cur}}{c}, 
 \label{eq3.29} 
\end{equation}
where $\Gamma_{e}$ is the particle Lorentz factor and $P_{IC}/c$ and $P_{cur}/c$ are the 
drag forces caused by IC scattering and curvature radiation, respectively.

Leptons, moving along field lines within the gap, upscatter the soft photons radiated from 
the inner region of the ADAF. The drag force due to IC emission (i.e., in units 
of erg~cm$^{-1}$) is defined by \citep[e.g.,][]{hir99}
\begin{eqnarray}
 \frac{P_{IC}}{c}=\int\limits_{E_{s}^{min}}^{m_{e}c^2/\Gamma_{e}}E_{\gamma}\,\sigma_{KN}\,
 \frac{dN_{s}}{dE_{s}}
 \, dE_{s}+ \quad \quad \quad \quad \nonumber \\ 
 +\int\limits_{m_{e}c^2/\Gamma_{e}}^{E_{s}^{max}} E_{\gamma}\,\sigma_{KN}\,
 \frac{dN_{s}}{dE_{s}}\, dE_{s}, \label{eq3.30} 
\end{eqnarray}
where $dN_{s}/dE_{s}$ is number density of the ADAF soft photons per unit energy\footnote{This
is estimated using $dN_{s}/dE_{s}=F_{\nu}/4\pi d^2chE_{s}$, where $F_{\nu}$ is the spectral 
flux of the considered ADAF in units of erg~s$^{-1}$ Hz$^{-1}$, and $E_{s}=h\nu$ is the soft 
photon energy. In the following this is evaluated for a sphere with radius $d =5 r_{g}$.} 
and $\sigma_{KN}$ is the total Klein-Nishina cross section \citep{ryb79},
\begin{eqnarray}
\sigma_{KN}(x)=
\frac{3}{4}\sigma_{\tau}\left\{\frac{1+x}{x^3}\left[\frac{2x(1+x)}{1+2x}-\right. 
                \right. \quad \quad \quad \quad \nonumber \\ 
\left. \left. \ln(1+2x)\right]+\frac{1}{2x}\ln(1+2x)-\frac{1+3x}{(1+2x)^2}\right\}, 
\label{eq3.31} 
\end{eqnarray}
where $x=E_{s}\Gamma_{e}/m_{e}c^2$ and $\sigma_{\tau}$ is the Thomson cross section. The 
transition from Thomson to the Klein-Nishina limit roughly occurs at energy $E_{s}^{t}
\approx m_{e}c^2/\Gamma_{e}$. If the initial soft photon has an energy smaller than this 
(i.e., $E_{s}<E_{s}^{t}$), then its post-collision $\gamma$-ray energy is on average 
$E_{\gamma}\approx\Gamma_{e}^2E_{s}$. On the other hand, for $E_{s}>E_{s}^{t}$ the 
upscattered photon energy is limited by the energy of the electron, $E_{\gamma}=
\Gamma_{e}m_{e}c^2$, in the Klein-Nishina limit.

We also consider that leptons can emit $\gamma$-ray curvature photons. The drag force due 
to curvature emission (i.e., in units erg~cm$^{-1}$) is \citep[e.g.,][]{rie11}
\begin{equation}
\frac{P_{cur}}{c}=\frac{2e^2}{3R_{c}^2}\,\Gamma_{e}^4. \label{eq3.32} 
\end{equation}
For the results shown below, a typical value for the curvature radius of $R_{C} \approx 
r_{g}$ has been assumed. In general, curvature losses become only relevant for very high 
Lorentz factors (typically above $\Gamma_e=10^{9.5}$), with inverse Compton usually 
providing the dominant loss channel. For accretion rates smaller than $\sim10^{-4}$, 
however, curvature losses become relevant at even lower Lorentz factors. 

Relation (\ref{eq3.29}) together with the expressions (\ref{eq3.30}) and (\ref{eq3.32}), 
provides the second equation of the system that describes the gap structure. It is worth 
commenting, at this point, on equation (\ref{eq3.30}). As can be seen, the dependence of 
the Lorentz factor is within the integrals as well, thereby complicating the numerical 
calculation. In order to reduce complexity, we thus approximate the Compton losses with 
a fifth-order polynomial function in the following.

\subsection{The lepton distribution} \label{subsec:03.03} 
The existence of leptons within the gap might be the result of more than one physical 
process. The primary particles, for example, could be injected via annihilation of ADAF 
MeV photons, or via diffusion \citep{lev11}. Here we explore the case where the pair 
cascade, which develops inside the gap, dominates the particle ($e^{\pm}$) densities and 
the structure of the gap. By definition, the total charge density within the gap must not 
be in excess (i.e., $\rho_{e}<\rho_{GJ}$). We consider that the pair cascade does not 
provide full screening everywhere, maintaining in such a way a stationary gap within the 
magnetosphere.

The distribution of electrons and positrons inside the gap can be found by means of the 
continuity equation \citep[e.g.,][]{hir98,hir99}. Assuming that the black hole magnetosphere 
is in steady state (i.e., $\partial/\partial t=0$), the continuity equation for both species 
($e^{\pm}$) is given by
\begin{equation}
\nabla\cdot\textbf{J}^{\pm}=\mathcal{S}^{\pm}, \label{eq3.33} 
\end{equation}
where $\textbf{J}^{\pm} = \rho^{\pm}\textbf{v}_{e}^{\pm}$ is the vector of current, with 
$\rho^{\pm}$ and $\textbf{v}_{e}^{\pm}$ the charge density and the velocity of positrons 
and electrons, respectively, and $\mathcal{S}^{\pm}$ is the source term explained below. 
The positive sign in (\ref{eq3.33}) refers to positrons, and the negative one to electrons.  

In the considered 1D approximation one then obtains for the radial distribution 
of positrons and electrons
\begin{equation}
 -\frac{d}{dr}\left[\rho^{+}c\left(1-\frac{1}{\Gamma_{e}^2}\right)^{\frac{1}{2}}\right]
 =\mathcal{S}^{+}, 
 \label{eq3.35} 
 \end{equation}
 \begin{equation} 
 \frac{d}{dr}\left[\rho^{-}c\left(1-\frac{1}{\Gamma_{e}^2}\right)^{\frac{1}{2}}\right]
 =\mathcal{S}^{-}. 
 \label{eq3.36} 
\end{equation}
As discussed before, the accelerated leptons emit gamma-rays owing to Compton upscattering 
of disk photons. The resulting high-energy photons are able to annihilate with soft ones, 
producing more pairs within the gap accelerator. Consequently, leptons coming from all 
generations are included in equations (\ref{eq3.35}) and (\ref{eq3.36}) by means of the 
source function $\mathcal{S}^{\pm}$. 

In order to estimate $\mathcal{S}^{\pm}$, let us consider the distribution of $\gamma$-ray 
photons $P_{\gamma}^{\pm}(r,E_{\gamma})$ (i.e., number of photons per unit volume per unit 
energy), where the ``$+$'' sign denotes photons which move outward and the ``$-$'' sign those 
moving toward the BH. For a given energy range (i.e., from $E_{\gamma}$ to
$E_{\gamma}+dE_{\gamma}$) the number of photons per unit volume is simply
$\left[P_{\gamma}^{+}(r,E_{\gamma})+P_{\gamma}^{-}(r,E_{\gamma})\right]dE_{\gamma}$. Since 
not all the photons will contribute efficiently, the number of photons needs to be multiplied 
by a corresponding coefficient. For pair production this coefficient is 
\begin{equation}
 \alpha_{p}(E_{\gamma})=\int\limits_{\frac{(m_{e}c^2)^2}{E_{\gamma}}}^{E_{s}^{max}} 
 \sigma_{p}\,\frac{dN_{s}}{dE_{s}}\,dE_{s}, \label{eq3.37} 
\end{equation}
where $\sigma_{p}$ is the  pair production cross section in a collision between two photons 
with energies $E_{s}$ and $E_{\gamma}$, and given by \citep{ber82}
\begin{eqnarray}
 \sigma_{p}=\frac{3}{16}\sigma_{\tau}(1-\beta_{*}^2)\left[(3-\beta_{*}^{4})
 \ln\left(\frac{1+\beta_{*}}{1-\beta_{*}}\right)-\right. \quad \nonumber \\
 \left. -2\beta_{*}(2-\beta_{*}^{2})\right], \label{eq3.38} 
\end{eqnarray}
where $\beta_{*}=\sqrt{1-m_{e}^2c^4/E_{s}E_{\gamma}}$. We note that for the numerical 
evaluation of the system only head-on photon collisions have been taken into account. 
For a given $E_{\gamma}$, the coefficient $\alpha_p$ is active only when the colliding 
soft photon has energy above the threshold, that is $E_{s}\geq (m_{e}c^2)^2/E_{\gamma}$. 
Eventually, the number density of particles per unit length that is injected into the gap due 
to photon-photon pair production is given $a_{p}(E_{\gamma})[P_{\gamma}^{+}(r,E_{\gamma})+
P_{\gamma}^{-}(r,E_{\gamma})]dE_{\gamma}$. Therefore, the total pair production rate 
(i.e., incoming charge density per unit time) becomes
\begin{equation}
 \mathcal{S}^{\pm}=
 \pm ec\int\limits_{0}^{\infty} a_{p}(E_{\gamma})[P_{\gamma}^{+}(r,E_{\gamma})
  +P_{\gamma}^{-}(r,E_{\gamma})]dE_{\gamma}\,. \label{eq3.39} 
\end{equation}
In principle, curvature photons also contribute to pair production. In the ADAF case, 
however, the number density of soft photons above the corresponding threshold for 
curvature photons is much smaller compared to that for IC, suggesting that 
curvature emission only makes a negligible contribution to the total pair production rate.

Adding equations (\ref{eq3.35}) and (\ref{eq3.36}) one finds
\begin{equation}
  \frac{d}{dr}
  \left[(\rho^{-}-\rho^{+})c\left(1-\frac{1}{\Gamma_{e}^2}\right)^{\frac{1}{2}}\right]=0. 
  \label{eq3.40} 
\end{equation} implying that the total current $J_0$ along a magnetic field line is constant, 
i.e.,
\begin{equation}
 J_0 = (\rho^{-}-\rho^{+})c\left(1-\frac{1}{\Gamma_{e}^2}\right)^{\frac{1}{2}}
 ={\mathrm{constant}}.\, 
 \label{eq3.41} 
\end{equation}
If we subtract, on the other hand, equations (\ref{eq3.35}) and (\ref{eq3.36}) we find
\begin{eqnarray}
  \frac{d}{dr}\left[(\rho^{+}+\rho^{-})c\left(1-\frac{1}{\Gamma_{e}^2}\right)^{\frac{1}{2}}\right]= 
  \quad \quad \quad \quad \quad \nonumber \\
  -2ec\int\limits_{0}^{\infty} a_{p}(P_{\gamma}^{+}+P_{\gamma}^{-})dE_{\gamma}.\label{eq3.42} 
\end{eqnarray}
Instead of equations (\ref{eq3.35}) and (\ref{eq3.36}), the relations (\ref{eq3.40}) and 
(\ref{eq3.42}) are added to the system that describes the structure of the gap. Below, we 
close the set of equations, giving the expressions for the distributions of $\gamma$-ray 
photons.

\subsection{The $\gamma$-ray photon distributions} \label{subsec:03.04}
The distribution of high energy photons for the (1D) gap accelerator in steady state is 
described by the (Boltzmann) transport equation
\begin{equation}
 \pm c\frac{d}{dr}P_{\gamma}^{\pm}(r,E_{\gamma})=\mathcal{N}^{\pm}\,,  \label{eq3.43} 
\end{equation}
where $\mathcal{N}^{\pm}$ represents the rate of change of the number density of 
photons per unit energy. This can be written as $\mathcal{N}^{\pm}=
(\mathcal{N}_{\rm gain}-\mathcal{N}_{\rm loss})^{\pm}$, 
where the term $\mathcal{N}_{\rm gain}$ represents $\gamma$-ray photons added to the 
system, while the term $\mathcal{N}_{\rm loss}$ represents photons that leave the system. 
We can easily express $\mathcal{N}_{\rm loss}$, since we have already defined the 
distribution of $\gamma$-ray photons $P_{\gamma}^{\pm}$ and the possibility for
$\gamma\gamma$-annihilation, i.e.,
\begin{equation}
 \mathcal{N}_{\rm loss}=c\, \alpha_{p}\, P_{\gamma}^{\pm}(r,E_{\gamma})\,. 
 \label{eq3.44} 
\end{equation}
We remind that the photon distributions $P_{\gamma}^{\pm}$ include not only the up-scattered 
photons, but the curvature ones as well. 

On the other hand, the photon population in equation (\ref{eq3.43}) also increases due to 
the $\gamma$-ray production taking place within the accelerating area. In particular, 
$\mathcal{N}_{\rm gain} = \mathcal{N}_{\rm gain}^{IC}+\mathcal{N}_{\rm gain}^{cur}$, since 
photons emitted by both, inverse Compton and curvature processes contribute to 
$\mathcal{N}_{\rm gain}$. In order to estimate $\mathcal{N}_{\rm gain}^{IC}$, consider the number 
density of particles, $n^{\pm}/ \approx \pm \rho^{\pm}/e$. Since not 
all the particles up-scatter soft photons with the same efficiency, we need to construct 
a relevant coefficient to determine the rate of scattered photons. This coefficient can 
be written as \citep{hir99}
\begin{eqnarray}
 \alpha_{IC}(E_{\gamma},\Gamma_{e})= 
 \quad \quad \quad \quad \quad \quad \quad \quad \quad \quad \quad \quad \quad \quad \nonumber \\ 
 \frac{1}{m_{e}c^2} \int\limits_{E_{s}^{min}}^{E_{s}^{max}}\sigma_{KN}\delta\left(\epsilon_{\gamma}
 -{\rm min}[\Gamma_{e}^2\epsilon_{s},\Gamma_{e}]\right)\frac{dN_{s}}{dE_{s}}dE_{s}, \label{eq3.46} 
\end{eqnarray}
where $\epsilon_{s}=E_{s}/m_{e}c^2$, $\epsilon_{\gamma}=E_{\gamma}/m_{e}c^2$ are the normalized 
(i.e., in units of the electron rest mass) energies of soft and $\gamma$-ray photon, respectively. 
Hence, one has
\begin{equation}
 \mathcal{N}_{\rm gain}^{IC}=\pm \alpha_{IC}\,\frac{\rho^{\pm}}{e m_{e}c^2}\,c\,
 \sqrt{1-\frac{1}{\Gamma_{e}^2}}\,.
 \label{eq3.47} 
 \end{equation}

The term $\mathcal{N}_{\rm gain}^{cur}$, on the other hand, can be expressed by considering 
the curvature power emitted by a single electron. We can approximate this using the synchrotron
formula and assuming that the relativistic electron moves along a field line with curvature
radius $R_{c}=\Gamma_{e}m_{e}c^2/(eB\sin\theta_{c})\approx r_{g}$. Accordingly, the emitted
spectral power (in units of erg~sec$^{-1}$~Hz$^{-1}$) can be written as \citep[e.g.,][]{ryb79} 
\begin{equation}
 p_{cur}=\frac{\sqrt{3}\,e^2}{r_{g}}\,\Gamma_{e} ~F\left(\frac{E_{\gamma}}{E_{c}}\right),
\label{eq3.48} 
\end{equation}
with $F(x)$ in equation (\ref{eq3.48}) given by 
\begin{equation}
 F(x)=x\int\limits_{x}^{\infty}K_{\frac{5}{3}}(z)dz\approx x^{0.3}e^{-x}, 
 \label{eq3.49} 
\end{equation}
where $K_{5/3}$ is the modified Bessel function of order of $5/3$, and $x=E_{\gamma}/E_{c}$. 
The critical value $E_{c}$ corresponds to the energy at which most of the emission takes 
place, i.e.,
\begin{equation}
 E_{c}=\frac{3}{4\pi}\,\frac{hc}{r_{g}}\,\Gamma_{e}^{3}. \label{eq3.50} 
\end{equation}
Dividing equation (\ref{eq3.48}) by $h\epsilon_{\gamma} m_{e}c^2$ and multiplying simultaneously 
with the number density of particles (i.e., $\pm\rho^{\pm}/e$), we obtain the total 
contribution of the curvature process to $\mathcal{N}_{\rm gain}$. This increased rate of the number 
density of photons per unit energy due to curvature radiation is
\begin{equation}
 \mathcal{N}_{\rm gain}^{cur}=\pm \alpha_{cur}\,c\, \frac{\rho^{\pm}}{e\, m_{e}c^2}, \label{eq3.51}  
\end{equation}
where the coefficient $\alpha_{cur}$ is given by
\begin{equation}
\alpha_{cur}(E_{\gamma},\Gamma_{e})=\frac{\sqrt{3}\,e^2}{h\, r_{g} \epsilon_{\gamma}c}\, 
\Gamma_{e}\,F\left(\frac{E_{\gamma}}{E_{c}}\right). \label{eq3.52}  
\end{equation}
Note that the terms $\mathcal{N}_{\rm gain}$ and $\mathcal{N}_{\rm losses}$ should be considered
with respect to the direction of particle motion.

Using the considerations above, we can formulate the expressions that describe the distribution 
of $\gamma$-ray photons within the gap accelerator. Substituting equations (\ref{eq3.44}), 
(\ref{eq3.47}) and (\ref{eq3.51}) into formula (\ref{eq3.43}), one finds
\begin{eqnarray}
 c\frac{dP_{\gamma}^{+}}{dr}=-a_{IC}\,\frac{\rho^{-}}{e m_{e}c^2}\,c\,
 \sqrt{1-\frac{1}{\Gamma_{e}^2}}- \quad \quad \quad \quad \nonumber \\
 \alpha_{cur}\,c\,\frac{\rho^{-}}{e m_{e}\,c^2}-c\,a_{p}\,P_{\gamma}^{+}, \label{eq3.53} 
 \end{eqnarray}
 \begin{eqnarray}
 c\frac{dP_{\gamma}^{-}}{dr}=-a_{IC}\,\frac{\rho^{+}}{e m_{e}c^2}\,c\,
 \sqrt{1-\frac{1}{\Gamma_{e}^2}}- \quad \quad \quad \quad  \nonumber \\
 \alpha_{cur}\,c\,\frac{\rho^{+}}{e m_{e}\,c^2}+c\,a_{p}\,P_{\gamma}^{-}. \label{eq3.54}
\end{eqnarray}
For the sake of clarity we mention again that electrons ($-\rho^-/e$) moving away from the BH
owing to the chosen field direction are responsible for the (outwardly moving) photon 
distribution $P_{\gamma}^{+}$ (see equation \ref{eq3.53}). Correspondingly, positrons 
($\rho^+/e$) that move toward the BH produce the photon distribution $P_{\gamma}^{-}$ 
(see equation \ref{eq3.54}).

Formulas (\ref{eq3.53}) and (\ref{eq3.54}) fully close the set of equations that determine the 
structure of the gap accelerator. To sum up, relations (\ref{eq3.20}), (\ref{eq3.29}), 
(\ref{eq3.40}), (\ref{eq3.42}), (\ref{eq3.53}) and (\ref{eq3.54}) form a well defined system of 
six equations with six unknown physical quantities (i.e.,  $\mathcal{E}_{||}^{r}$, $\Gamma_{e}$,
$\rho^{+}$, $\rho^{-}$, $P_{\gamma}^{+}$ and $P_{\gamma}^{-}$). 

\vspace{0.1cm}
\section{Normalization of the system} \label{sec:04}

Aiming to facilitate the numerical manipulation, we normalize and write the system of equations 
without physical units. Accordingly, lengths are expressed as $\xi=r/r_{g}$ and charge densities 
as $\rho_{*}^{\pm}=\rho^{\pm}/\rho_{c}$, where
\begin{equation}
 \rho_{c}=\frac{\Omega^{F}B_{H}}{2\pi c}\approx 2.69\times 10^{-11} M_{9}^{-3/2}\dot{m}^{1/2}, 
 \label{eq4.01} 
\end{equation} (units: statC cm$^{-3}$), noting that $\Omega^{F}=\Omega^{H}/2$, with 
$\Omega^{H}=\alpha_{s}c^3/2GMr_{H}$ the angular velocity of the black hole, and
$B_{H}=10^5\,\dot{m}^{1/2}M_{9}^{-1/2}$ G the magnetic field strength near the horizon 
\citep{kats18}. 

Gauss' law (\ref{eq3.20}) thus becomes
\begin{equation}
  \frac{d}{d\xi}\left(\frac{\mathcal{E}_{||}^{*r}}{\alpha_{l}}\right)
  =\rho_{*}^{+}+\rho_{*}^{-}-\rho_{GJ}^{*}
  -A^{*}\left(\frac{\mathcal{E}_{||}^{*r}}{\alpha_{l}}\right),            
  \label{eq4.02} 
\end{equation}
where $\rho_{GJ}^{*}=\rho_{GJ}/\rho_{c}$, $A^{*}=r_{g}(1/\sqrt{|\gamma|})
(d\sqrt{|\gamma|}/dr)$ and $\mathcal{E}_{||}^{*r}=\mathcal{E}_{||}^{r}/
4\pi r_{g}\rho_{c}$ is the normalized parallel electric field component 
(i.e., the contravariant one). 

For the equation of motion (\ref{eq3.29}) one finds
\begin{equation}
 \frac{d\Gamma_{e}}{d\xi}=-\mathcal{C}_{1}\mathcal{E}_{||}^{*r}
 -\mathcal{C}_{2}\mathcal{F}(\Gamma_{e})-\mathcal{C}_{3}\Gamma_{e}^4,   \label{eq4.05} 
\end{equation}
where the non-dimensional quantities $\mathcal{C}_{1}$ and $\mathcal{C}_{3}$ are given by
\begin{equation}
 \mathcal{C}_{1}=\frac{4\pi\, e\,r_{g}^2\,\rho_{c}}{m_{e}\,c^2}
 \approx 4.32\times 10^{15}M_{9}^{1/2}\dot{m}^{1/2}, \label{eq4.06} 
\end{equation}
and
\begin{equation}
 \mathcal{C}_{3}=\frac{2\,e^2}{3\,r_{g}\,m_{e}\,c^2}
 \approx 0.13 \times 10^{-26}M_{9}^{-1}\,.\label{eq4.07} 
\end{equation}
The Compton term in equation (\ref{eq4.05}) is $\mathcal{C}_{2}\mathcal{F}(\Gamma_{e})=
(r_{g}/m_{e}c^2)(P_{IC}/c)$. 

Based on the continuity equation for the leptons (\ref{eq3.40}) one obtains
\begin{equation}
  \frac{d}{d\xi}\left[(\rho_{*}^{-}-\rho_{*}^{+})
  \left(1-\frac{1}{\Gamma_{e}^2}\right)^{\frac{1}{2}}\right]=0, \label{eq4.08} 
\end{equation}
with
\begin{equation}
 (\rho_{*}^{-}-\rho_{*}^{+})\left(1-\frac{1}{\Gamma_{e}^2}\right)^{\frac{1}{2}}
 =\frac{J_{o}}{c\,\rho_{c}}=J_{o}^{*}\,, \label{eq4.09} 
\end{equation}
where the constant parameter $J_{o}^{*}$ is the dimensionless current density which 
corresponds to the global magnetospheric current. Note that $J_{o}^{*}$ is normalized 
via $\rho_{c}$ of equation (\ref{eq4.01}) and not via the relativistic Goldreich-Julian 
charge density that varies with $\xi$.

In addition equation (\ref{eq3.42}), which also describes the lepton population within 
the gap accelerator, becomes
\begin{eqnarray}
  \frac{d}{d\xi}\left[(\rho_{*}^{+}+\rho_{*}^{-})
  \left(1-\frac{1}{\Gamma_{e}^2}\right)^{\frac{1}{2}}\right]=  
  \quad \quad \quad \quad \quad  \nonumber \\
  -2\int\limits_{0}^{\infty} a_{p}^{*}(P_{\gamma^{*}}^{+}
  +P_{\gamma^{*}}^{-})d\epsilon_{\gamma}\,, \label{eq4.10} 
\end{eqnarray}
where $a_{p}^{*}=r_{g}\,a_{p}$ and $P_{\gamma^{*}}^{\pm}=(e\,m_{e}c^2/\rho_{c})\,P_{\gamma}^{\pm}$ 
represent the normalized outgoing/incoming $\gamma$-ray photons. Equation (\ref{eq4.10}) captures 
the information for the lepton distribution which is injected into the gap due to $\gamma$-ray 
photon annihilation. Following \citet{hir98} we approximate the integral of equation (\ref{eq4.10}) 
for numerical reasons by a summation, dividing the $\gamma$-ray energy band into many ($m$) finite 
energy bins (we typically apply $m=80$ energy bins). Hence, we eventually obtain
 \begin{eqnarray}
  \frac{d}{d\xi}\left[(\rho_{*}^{+}+\rho_{*}^{-})
  \left(1-\frac{1}{\Gamma_{e}^2}\right)^{\frac{1}{2}}\right]
  = \quad \quad \quad \quad \quad \quad  \nonumber \\
  -2\sum\limits_{i=1}^{m} a_{p,i}^{*}(\mathcal{P}_{*,i}^{+}+\mathcal{P}_{*,i}^{-})\,, 
  \label{eq4.14} 
\end{eqnarray} where
\begin{equation}
 a_{p,i}^{*} \approx a_{p}^{*}\left(\frac{\epsilon_{\gamma}^{(i-1)}
 +\epsilon_{\gamma}^{(i)}}{2}\right),\quad  \mathcal{P}_{*,i}^{\pm}
 =\int\limits_{\epsilon_{\gamma}^{(i-1)}}^{\epsilon_{\gamma}^{(i)}}P_{\gamma*}^{\pm}
 \,d\epsilon_{\gamma}\,.  \label{eq4.13} 
\end{equation}

For the outcoming/incoming distribution of $\gamma$-ray photons, equations (\ref{eq3.53}) 
and (\ref{eq3.54}), one finds
\begin{equation}
 \pm \frac{dP_{\gamma^{*}}^{\pm}}{d\xi}=\mp \alpha_{IC}^{*}\,\rho_{*}^{\mp}\,
 \sqrt{1-\frac{1}{\Gamma_{e}^2}}\mp\alpha_{cur}^{*}\,
 \rho_{*}^{\mp}-a_{p}^{*}\,P_{\gamma^{*}}^{\pm}, \label{eq4.15} 
\end{equation}
with $\alpha_{IC}^{*}=r_g\, \alpha_{IC}$ and $\alpha_{cur}^{*}=r_g\, \alpha_{cur}$. 
Integrating this relation over energy interval, and using expression (\ref{eq4.13}) we can 
write
\begin{eqnarray}
\frac{d\mathcal{P}_{*,i}^{+}}{d\xi}
=-\alpha_{IC,i}^{*}\,\rho_{*}^{-}\,\sqrt{1-\frac{1}{\Gamma_{e}^2}}
- \quad \quad \quad \quad \quad  \nonumber \\
-\alpha_{cur,i}^{*}\,\rho_{*}^{-}-a_{p,i}^{*}\,\mathcal{P}_{*,i}^{+},\label{eq4.17}
\end{eqnarray}
\begin{eqnarray}
\frac{d\mathcal{P}_{*,i}^{-}}{d\xi}
=-\alpha_{IC,i}^{*}\,\rho_{*}^{+}\,\sqrt{1-\frac{1}{\Gamma_{e}^2}}
-  \quad \quad \quad \quad \quad \quad\nonumber \\
-\alpha_{cur,i}^{*}\,\rho_{*}^{+}+a_{p,i}^{*}\,\mathcal{P}_{*,i}^{-}, \quad  \label{eq4.18}
\end{eqnarray}
with coefficients $a_{IC,i}^{*}$ and $a_{cur,i}^{*}$ given by
\begin{equation}
 a_{IC,i}^{*}
 =\int\limits_{\epsilon_{\gamma}^{(i-1)}}^{\epsilon_{\gamma}^{(i)}}a_{IC}^{*}\,
 d\epsilon_{\gamma}, \quad  a_{cur,i}^{*}
 =\int\limits_{\epsilon_{\gamma}^{(i-1)}}^{\epsilon_{\gamma}^{(i)}}a_{cur}^{*}
 \,d\epsilon_{\gamma}.  
 \label{eq4.19} 
\end{equation}

Hence, relations (\ref{eq4.02}), (\ref{eq4.05}), (\ref{eq4.08}), (\ref{eq4.14}), 
(\ref{eq4.17}) and (\ref{eq4.18}) form the normalized system of ``$4+2m$'' equations 
that govern the physics of the gap accelerator. Imposing suitable boundary conditions 
we then integrate the system numerically and determine the structure of the gap, that 
is the radial distributions of $\mathcal{E}_{||}^{*r}$, $\Gamma_{e}$, $\rho_{*}^{+}$, 
$\rho_{*}^{-}$, $\mathcal{P}_{*,i}^{+}$ and $\mathcal{P}_{*,i}^{-}$.

\section{The boundary conditions} \label{sec:05}
The aforementioned system of equations constitutes a boundary value problem, 
since conditions that reflect the gap physics have to be satisfied at the inner 
and the outermost gap positions. We use $\xi_{1}$ to denote the inner boundary of 
the gap, and $\xi_{2}$ for the outer one in the following.

Typical boundary conditions are discussed in, e.g., \citet{hir98} and \citet{lev17}. 
Accordingly, we impose that the parallel component of the electric field vanishes at 
both boundaries. Hence, we have
\begin{equation}
  \left. \mathcal{E}_{||}^{*r}\right|_{\xi_{1}}=0, 
  \qquad \left.\mathcal{E}_{||}^{*r}\right|_{\xi_{2}}=0. \label{eq5.01} 
\end{equation}
The emergence of a parallel electric field, which is the result of a charge deficit 
in the region, is terminated at $\xi_{1,2}$ ensuring force-freeness beyond the gap 
boundaries. Therefore, particle acceleration is no longer possible at the boundaries, 
so that 
\begin{equation}
  \left.\Gamma_{e}\right|_{\xi_{1}}=1, \qquad \left.\Gamma_{e}\right|_{\xi_{2}}=1. 
  \label{eq5.02} 
\end{equation} The numerical solutions are, however, not very sensitive to this 
condition.

Using equation (\ref{eq4.09}) and taking into account that the electric field directs 
positrons toward the event horizon and electrons outward, an idealized situation has 
been previously considered \citep{hir98} where
\begin{equation}
 \left.\rho_{*}^{-}\right|_{\xi_{1}}=0, \qquad
 \left.\rho_{*}^{+}\right|_{\xi_{1}}=-\frac{J_{o}^{*}}{\sqrt{1-\frac{1}{\Gamma_{e}^2}}}, 
 \label{eq5.03} 
\end{equation}
at the inner boundary position $\xi_{1}$, and
\begin{equation}
 \left.\rho_{*}^{+}\right|_{\xi_{2}}=0, \qquad
 \left.\rho_{*}^{-}\right|_{\xi_{2}}=\frac{J_{o}^{*}}{\sqrt{1-\frac{1}{\Gamma_{e}^2}}}, 
 \label{eq5.04} 
\end{equation}
at the outer boundary position $\xi_{2}$\footnote{We note that the global magnetospheric 
current $J_{0}^{*}$ takes a negative value in our convention (where the electric field points 
toward the BH), resulting in positive $\left.\rho_{*}^{+}\right|_{\xi_{1}}$ and 
negative $\left.\rho_{*}^{-}\right|_{\xi_{2}}$ charge densities in equations (\ref{eq5.03}) 
and (\ref{eq5.04}), respectively.}. For such a choice, only positrons are present at the 
inner boundary, and only electrons at the outer one. Since this is generally somewhat 
artificial, we relax conditions (\ref{eq5.03}) and (\ref{eq5.04}) in our study, allowing 
for the possibility of some charge injection at the gap boundaries.  

Finally, for a gap accelerator assumed to be isolated from any other source of gamma-ray 
photons in the close vicinity of the black hole, one can further explore the case 
\citep[e.g.,][]{hir98,hir99,hir17}
\begin{equation}
  \left.\mathcal{P}_{*,i}^{+}\right|_{\xi_{1}}=0, 
  \qquad \left.\mathcal{P}_{*,i}^{-}\right|_{\xi_{2}}=0, \label{eq5.05} 
\end{equation} where high-energy $\gamma$-ray photons are not injected through 
the gap boundaries. However, even if particle acceleration terminates beyond the 
boundaries, the electromagnetic cascade can remain active for many gravitational radii. 
Accordingly, we may expect that some part of the high-energy photons produced outside 
the gap to get injected into it, at least through the outer boundary $\xi_{2}$. 
Therefore, we relax the condition (\ref{eq5.05}), and accept any choice of photon 
values $\left.\mathcal{P}_{*,i}^{\pm}\right|_{\xi_{1},\xi_{2}}$ that ultimately 
results in a charge density lower than the Goldreich-Julian one along the whole 
extension of the gap.

In short, we integrate the set of equations imposing conditions (\ref{eq5.01}), 
(\ref{eq5.02}) and demanding the resultant amount of charges to be less than the 
Goldreich-Julian charge density (i.e., $|\rho_{e}|=|\rho_{*}^{+}+\rho_{*}^{-}|\leq 
|\rho_{GJ}^{*}|$), irrespectively of whether conditions (\ref{eq5.03}), (\ref{eq5.04}) 
and (\ref{eq5.05}) are fully satisfied.

\section{Existence of steady gap solutions} \label{sec:06}
As described above, the Goldreich-Julian charge density changes sign across the null 
surface (from positive to negative, on moving outward), where $\Omega^{F}=\omega$ and 
$\rho_{GJ}=0$. The real charge distribution resulting from the integration of the 
system is around the Goldreich-Julian one, and its divergence from it gives the parallel 
electric field (equation [\ref{eq3.17}]). As the strength of the electric field is 
negative in our convention (i.e., it points toward the BH), we qualitatively 
expect that it starts to decrease from zero at the boundary $\xi_{1}$, then reaches a 
minimum at a certain distance, in which $\rho_{e}\approx \rho_{GJ}$, before it increases 
again up to zero at the boundary $\xi_{2}$. Hence, Gauss's law at the inner boundary 
becomes
\begin{equation}
  \left.\frac{d}{d\xi}\left(\frac{\mathcal{E}_{||}^{*r}}{\alpha_{l}}\right)\right|_{\xi_{1}}
  =\left.\rho_{*}^{+}+\rho_{*}^{-}-\rho_{GJ}^{*}\,\right|_{\xi_{1}}\leq 0,            
  \label{eq6.01} 
\end{equation}
while at the outer boundary
\begin{equation}
  \left.\frac{d}{d\xi}\left(\frac{\mathcal{E}_{||}^{*r}}{\alpha_{l}}\right)\right|_{\xi_{2}}
  =\left.\rho_{*}^{+}+\rho_{*}^{-}-\rho_{GJ}^{*}\,\right|_{\xi_{2}}\geq 0,            
  \label{eq6.02} 
\end{equation}
where $A^{*}(\mathcal{E}_{||}^{*r}/\alpha_{l})=0$ in equation (\ref{eq4.02}), using 
condition (\ref{eq5.01}). Both formulas (\ref{eq6.01}) and (\ref{eq6.02}) ensure that 
the charge density at the boundaries is not super-critical, i.e. $|\rho_{e}|\leq
|\rho_{GJ}|$ applies, since the Goldreich-Julian density is positive at $\xi_{1}$ and 
negative at $\xi_{2}$. For $|\rho_{e}|=|\rho_{GJ}|$ "brim" boundary solutions of the 
electric field are found \citep{hir98,hir17}.

\begin{figure}[ht!]
\hspace{-0.3cm}
\includegraphics[width=0.48\textwidth,height=7.3cm]{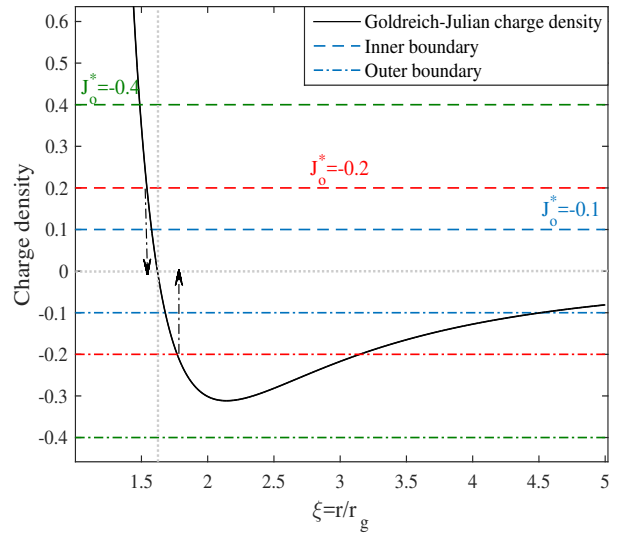}
\caption{Illustration of the Goldreich-Julian charge density (black solid line), 
the left-hand side of equation (\ref{eq6.05}) (dashed lines) and the left-hand side 
of equation (\ref{eq6.08}) (dashed-dotted lines) for current values of $J_{o}^{*}=-0.1$
(blue), $-0.2$ (red), and $-0.4$ (green), respectively. \label{fig:6.1}}
\end{figure}

Assuming that electron injection can occur across the boundary $\xi_{1}$, we express 
this as a fraction of the positron charge density,
\begin{equation}
  \left.\rho_{*}^{-}\right|_{\xi_{1}}=
  -n_{e} \left.\rho_{*}^{+}\right|_{\xi_{1}},  
  \label{eq6.03} 
\end{equation}
where $0\leq n_{e}<1$. Using equations (\ref{eq6.01}, \ref{eq4.09}) and (\ref{eq6.03}) 
one obtains
\begin{equation}
 \left.\left(\frac{n_{e}-1}{n_{e}+1}\right)
 \frac{J_o^{*}}{\sqrt{1-\frac{1}{\Gamma_{e}^2}}}\right|_{\xi_{1}}
 \leq  \left.\rho_{GJ}^{*}\right|_{\xi_{1}}. 
 \label{eq6.04} 
\end{equation} 
Equation (\ref{eq6.04}) implies that the inner boundary $\xi_{1}$ is constrained by 
the value of the current $J_{o}^{*}$ and the amount of injected electrons $n_{e}$. Assuming 
$n_e=0$ for convenience and keeping the equality in relation (\ref{eq6.04}), 
the innermost boundary\footnote{The innermost boundary $\xi_{1}$ relative to the 
radial distance where $\rho_{GJ}$ becomes zero.} can be estimated via
\begin{equation}
 \left.-\frac{J_o^{*}}{\sqrt{1-\frac{1}{\Gamma_{e}^2}}}\right|_{\xi_{1}}= \left.\rho_{GJ}^{*}\right|_{\xi_{1}}\,.    
 \label{eq6.05} 
\end{equation}
Similarly, using relation (\ref{eq6.02}) we can write
\begin{equation}
 \left.\left(\frac{1-n_{p}}{n_{p}+1}\right)
 \frac{J_o^{*}}{\sqrt{1-\frac{1}{\Gamma_{e}^2}}}\right|_{\xi_{2}}
 \geq  \left.\rho_{GJ}^{*}\right|_{\xi_{2}}\,,        
 \label{eq6.06} 
\end{equation}
where $0\leq n_{p}<1$ is the fraction of positrons injected across the outer 
boundary $\xi_{2}$,
\begin{equation}
  \left.\rho_{*}^{+}\right|_{\xi_{2}}=-n_{p}\,\left.\rho_{*}^{-}\right|_{\xi_{2}}.       
  \label{eq6.07} 
\end{equation}
Assuming again $n_{p}=0$ for convenience, the radial range of the outer boundary 
$\xi_{2}$ can be estimated via 
\begin{equation}
 \left.\frac{J_o^{*}}{\sqrt{1-\frac{1}{\Gamma_{e}^2}}}\right|_{\xi_{2}}=
 \left.\rho_{GJ}^{*}\right|_{\xi_{2}}.  
 \label{eq6.08} 
\end{equation}
Figure (\ref{fig:6.1}) shows the radial distribution of the Goldreich-Julian charge 
density (black solid line), the left-hand side of equation (\ref{eq6.05}) (dashed 
lines) and the left-hand side of equation (\ref{eq6.08}) (dashed-dotted lines) for 
three different values of the current, i.e., $J_{o}^{*}=-0.1,-0.2,-0.4$.
The intersection points of the horizontal (current) lines with the Goldreich-Julian 
charge density determine the innermost boundary $\xi_{1}$ and the range of the outer 
boundary $\xi_{2}$. Since we investigate gaps across the null surface, we always 
require the boundary $\xi_{1}$ to be below the radius at which $\rho_{\rm GJ}$ 
becomes zero, and the boundary $\xi_{2}$ to be above it. In general, inequalities 
(\ref{eq6.04}) and (\ref{eq6.06}) apply to each possible $\xi_{1}$ and $\xi_{2}$. 

Hence, numerically the position of the inner gap boundary is constrained to be
within the radial interval from $r_{H}$ up to $\xi_{1}^{max}$ (i.e., see the 
arrow pointing downward), while the outer boundary ranges from $\xi_{2}^{min}$ 
(see the arrow pointing upward) up to $\xi_{2}^{max}$ (see the 
second intersection point, for instance, of the red dashed-dotted line).
Practically, we expect the boundary $\xi_{2}$ to be closer to $\xi_{2}^{min}$ 
than to $\xi_{2}^{max}$ (the upper limit $\xi_{2}^{max}$ characterizes extended 
gaps for which the environment is inefficient to sustain a steady electromagnetic
cascade). Accordingly, the higher the current value, the larger the gap extension 
for a given accretion rate (e.g., compare the intersection points for $J_{o}^{*}
=-0.1$ and $J_{o}^{*}=-0.2$). No abrupt change in the gap width is, however, 
expected for variations of the current, due to the rather smooth decrease of 
the Goldreich-Julian charge density around the null surface. 

Note that figure (\ref{fig:6.1}) indicates that for a current value $J_{o}^{*}=
-0.4$ the outer boundary $\xi_{2}$ cannot be properly defined. Hence, no steady 
gap solution exists beyond a certain current value. This agrees with similar 
findings by \citet{lev17}, according to which steady gap solutions can only 
exist under rather restrictive conditions. We do emphasize, however, that our 
result depends on the applicability of conditions (\ref{eq5.03}) and 
(\ref{eq5.04}).      

\begin{figure}[ht!]
\includegraphics[width=0.48\textwidth,height=7.4cm]{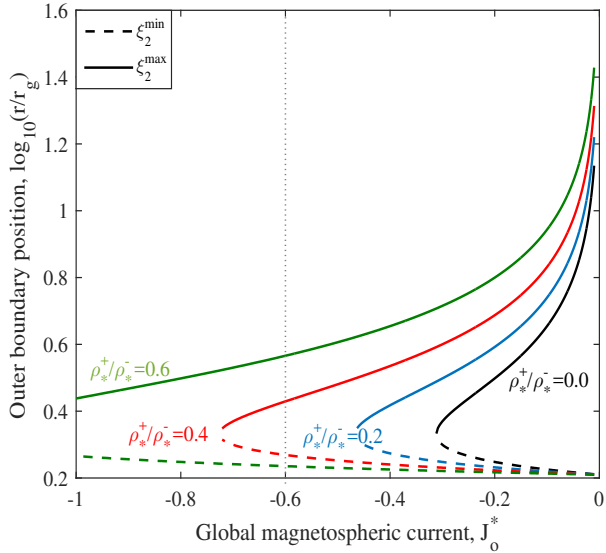}
\caption{Graphical illustration of the possible outer boundaries $\xi_{2}$ as 
function of the global magnetospheric current $J_{o}^{*}$ for four different 
fractions of positron injection: $\rho_{*}^{+}/\rho_{*}^{-}=0.0,0.2,0.4$ and 
$0.6$, given in black, blue, red and green colors, respectively. \label{fig:6.2}}
\end{figure}

In principle, the existence of an outer gap boundary $\xi_{2}$, and thus the 
existence of a steady gap solution in (1D), depends on the global magnetospheric 
current $J_{o}^{*}$ as well as on the positron fraction $n_{p}$ at the boundary. In 
figure (\ref{fig:6.2}), the radial range of the possible boundary $\xi_{2}$ 
is illustrated as function of the global magnetospheric current $J_{o}^{*}$ for 
four different values of positron injection (i.e., $n_{p}=0.0,0.2,0.4$ and $0.6$)
The dashed lines represent $\xi_{2}^{\rm min}$ and the solid lines  $\xi_{2}^{\rm 
max}$.
Evidently, we are able to define the radial range $\xi_{2}$ for a given value of 
the current only if the positron injection is sufficiently large (i.e., see for 
the dotted grey line). For instance, a steady gap solution cannot be found when
$n_{p}=\rho_{*}^{+}/\rho_{*}^{-}=0$ and $J_{o}^{*}<-0.3$. On the other hand, 
steady (1D) gaps might be sustainable for $J_{o}^{*}<-0.5$ if we relax condition 
(\ref{eq5.04}) and permit the injection of positrons at the outer gap boundary. 
We note that incorporating a (2D) electrodynamic structure may further help to
relax the constraints on steady gaps \citep[cf.][]{hir18}.

\section{Numerical Method} \label{sec:07}
In order to solve the system of equations, a shooting method \citep{pre07} is
applied. Specifically, we start integrating the equations from $\xi_{1}^{\rm max}$ 
up to a candidate $\xi_{2}$ (i.e., for a given accretion rate $\dot{m}$ and 
global current $J_{0}^{*}$), implementing the conditions at the inner boundary 
as initial ones. Then, we check whether the boundary conditions at $\xi_{2}$ are 
satisfied. Since we have relaxed conditions (\ref{eq5.03}), (\ref{eq5.04}) and 
(\ref{eq5.05}) in our approach, we iterate the integration changing the charge 
and photon injection. As discussed above, we require that a proper solution 
satisfies relations (\ref{eq5.01}) and (\ref{eq5.02}) at both boundaries and 
that the condition $|\rho_{e}|\leq |\rho_{GJ}^{*}|$ is ensured along the gap 
dimension. If a solution cannot be achieved for any choice of charge and photon
injection, we change $\xi_{1}$ moving toward the horizon and then integrate  
the set of equations again. If no solution is found through all possible 
$\xi_{1}$, we change the value of the current and subsequently the accretion 
rate until a gap solution can be found.

In figure (\ref{fig:7.1}) below, an example of the Goldreich-Julian charge 
density $\rho_{GJ}^{*}$ (blue line), the full expression $\rho_{GJ}^{*} +
A^{*}(\mathcal{E}_{||}^{*r}/\alpha_{l})$ of equation (\ref{eq4.02}) 
(dashed gray line), and a proper solution (black line) following consecutive 
numerical integrations (gray lines) are shown. As it can be seen, the full 
expression does not deviate much from $\rho_{GJ}^{*}$. The point where the 
Goldreich-Julian charge distribution becomes zero (hereafter, null point) is 
indicated by the intersection of the dashed black lines.
Imposing that the charge density remains less than the Goldreich-Julian 
charge distribution, we require our solution to pass through the null point, 
i.e., $\left.\rho_{e}\right|_{\xi_{null}} =\left.\rho_{GJ}\right|_{\xi_{null}}
=0$. This choice significantly reduces s the number of acceptable gap solutions. 
If, on the other hand, the charge density is not fixed relative to the null 
point, the resultant gaps would locally reveal a charge density higher than 
the Goldreich-Julian one \citep[cf.,][]{hirpu16,lev17}. The corresponding solutions 
tend to under/overestimate the gap width 
depending on the position of the electric field extremum, i.e., the gap size 
is underestimated when the minimum of the parallel electric field component 
occurs before the null point, and overestimated in the case where the minimum 
occurs beyond it. In order to take this into account, we consider that each steady 
gap realization should maintain a charge density below or equal to the 
Goldreich-Julian one. This is motivated by the fact that a possible 
surplus of charges, with their inherent tendency of adjustment to the 
critical value, is likely to cause dynamical oscillations to the gap, 
making its stability rather unlikely \citep{lev18}.\\ \\

\section{Solutions of the gap structure}\label{sec:08}
In the following subsections, we present solutions of the gap structure,
namely, the radial distribution of the physical quantities (e.g., the 
parallel electric field $\mathcal{E}_{||}^{*r}$, the particle Lorentz 
factor $\Gamma_e$, the charge density $\rho_e$ and the $\gamma$-ray 
photon spectrum) as obtained by solving 
the system of equations. In order to study the physics of the mechanism, 
explore its limits and compare with observations, we explore gap solutions 
for different values of the accretion rate and global magnetospheric current.
\begin{figure}[htb]
\includegraphics[width=0.46\textwidth,height=7.4cm]{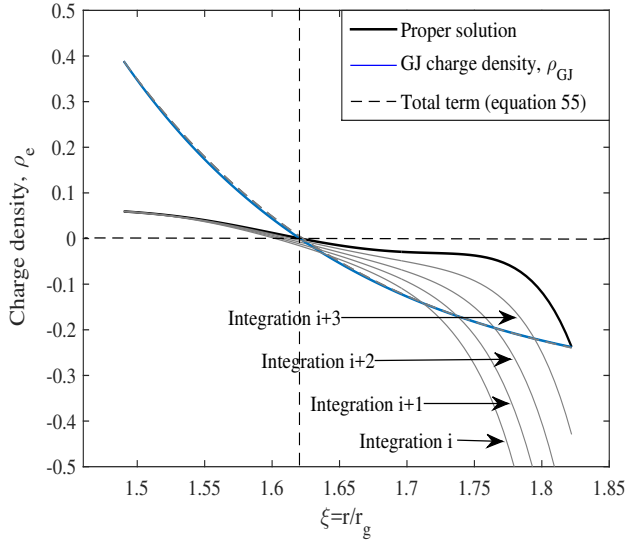}
\caption{Graphical illustration of the numerical method followed to find 
solutions of the gap structure. The blue curve represents the Goldreich-Julian 
charge density and the solid black curve the charge distribution, that results 
from several consecutive numerical integrations (gray curves).\label{fig:7.1}}
\end{figure}
\vspace{+0.3cm}
\begin{deluxetable}{cccc}[htb]
\tablenum{1}
\tablecaption{Gap properties for fixed accretion rate \label{tab_1}}
\tablewidth{0pt}
\tablehead{
\colhead{Global Current} & \colhead{Gap Size} & \colhead{Voltage Drop} 
          & \colhead{Gap Power}  \\
\colhead{$J_{o}^{*}=J_{o}/c\,\rho_{c}$} & \colhead{$h/r_{g}$} 
          & \colhead{$\times 10^{16}$ V} & \colhead{$\times 10^{40}\,erg\,s^{-1}$} 
}
\decimalcolnumbers
\startdata
$-0.005$   & $0.2550$ & $5.1$ & $0.1$ \\
$-0.157$   & $0.2879$ & $6.3$ & $2.2$ \\
$-0.29725$ & $0.3321$ & $7.3$ & $4.9$ \\ 
\enddata
\tablecomments{Results for the gap extension, the associated voltage drop, and 
total gap power for a fixed accretion rate of $\dot{m}=10^{-5.0}$ and a BH
with $M_9=1$ and $a_s^*=1$.}.
\end{deluxetable}

\subsection{Solutions for fixed accretion rate} \label{subsec:08.01}

\begin{figure*}[ht!]
\plottwo{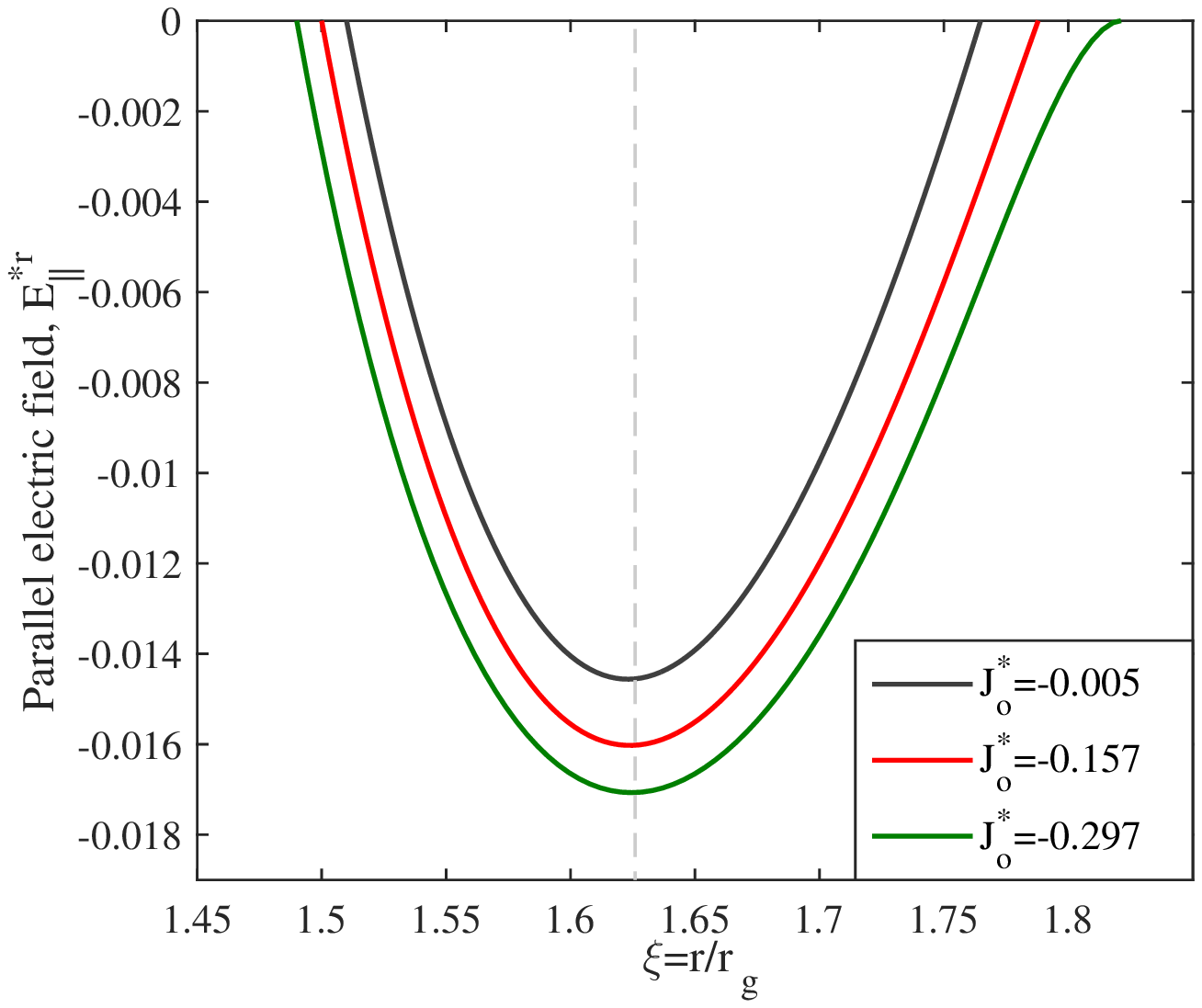}{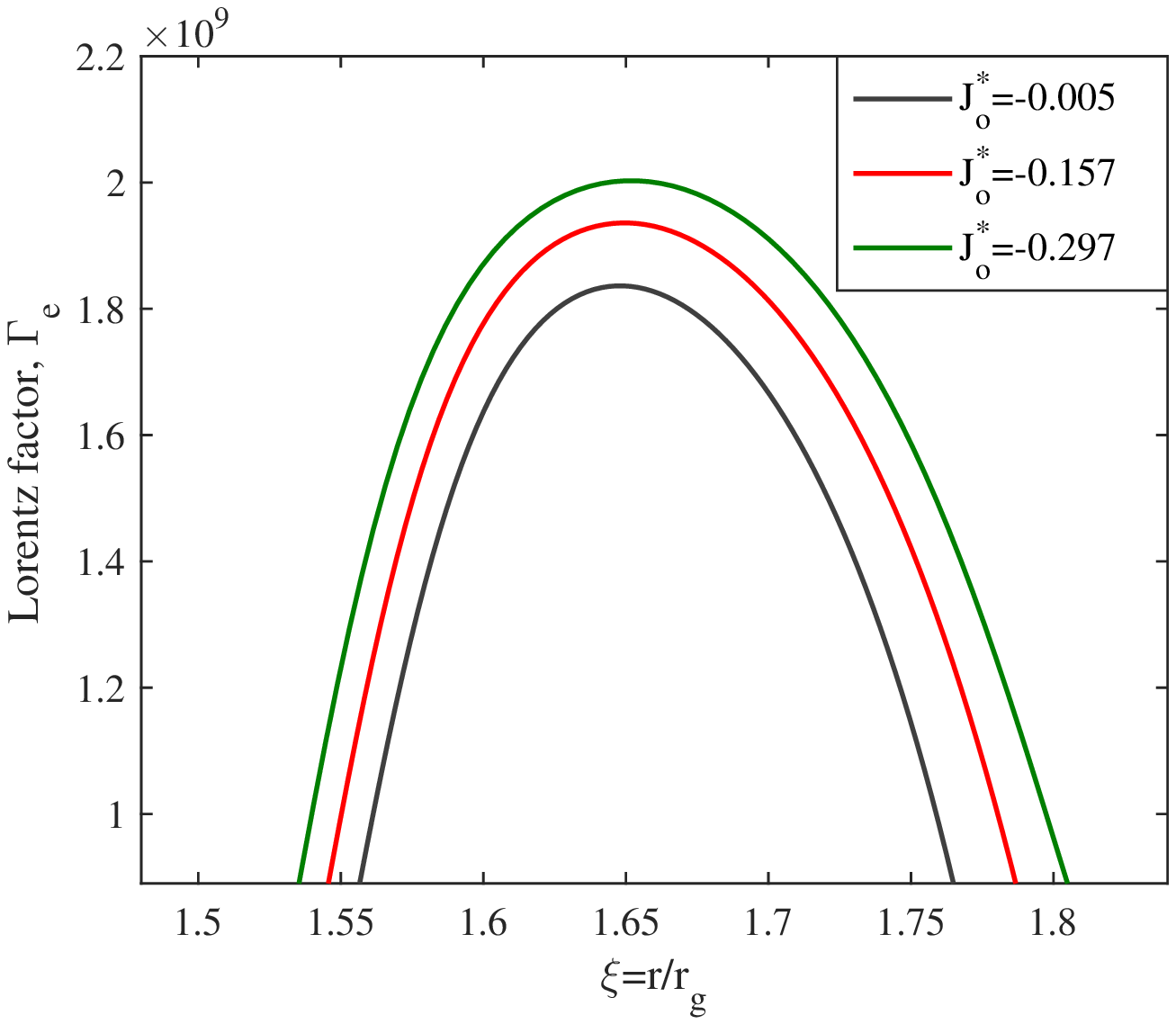}
\caption{Left: The normalized distribution of the parallel electric field 
component $\mathcal{E}_{||}^{*r}$ shown for current values $J_{o}^{*}=-0.005$ 
(black line), $J_{o}^{*}=-0.157$ (red line) and $J_{o}^{*}=-0.297$ (green line). 
Right: The corresponding Lorentz factor distribution $\Gamma_{e}$ of the 
particles.} \label{fig:8.1}
\vspace{+0.5cm}
\end{figure*}

Figure~(\ref{fig:8.1}) presents gap solutions for a fixed
accretion rate of $\dot{m}=10^{-5.0}$ and three different values of the current 
parameter, namely, $J_{o}^{*}=-0.005$, $J_{o}^{*}=-0.157$ and $J_{o}^{*}=-0.297$. 
A fast rotating ($\alpha_{s}^{*}=1.0$) supermassive ($M=10^{9}M_{\odot}$) BH
and a field line inclination $\theta=30^{o}$ have been assumed throughout.

The left panel of Figure~\ref{fig:8.1} shows that the gap extension increases as the 
amount of the global magnetospheric current increases. Roughly speaking, we 
obtain gap sizes smaller than $1/3$ of the gravitational radius for the 
parameters chosen here (see table~\ref{tab_1} for details). The electric 
field reaches its extremum at the null point as indicated by the dashed gray 
line. Figure (\ref{fig:8.1}, right) reveals that maximum Lorentz 
factors ($\Gamma_{e}\sim 10^9$) are achieved slightly beyond the minimum of the 
electric field. 
As can be seen, no dramatic changes in voltage drop or particle Lorentz factors 
are obtained for the considered current values. The resultant 
gap width here is essentially determined by the accretion rate, and only a weakly
dependent on the considered global current.

In table~\ref{tab_1} and in the following, the voltage drop is calculated by integrating the 
electric field, i.e., $\Delta V_{gap}=-\int_{\xi_{1}}^{\xi_{2}}r_{g}\,\mathcal{E}_{||}^{\hat{r}}\,d\xi$, 
while the gap power, $L_{gap} \propto J_0\,\Delta V_{gap}$, is estimated by the relation 
$L_{gap}=\int_{\xi_{1}}^{\xi_{2}}8\pi^2r_{g}\left(\frac{\rho_{e}}{e}\right)
\left(e\,\frac{dV_{gap}}{dr}\,c\right)\frac{\widetilde{\omega}\,\rho^2}{\sqrt{\Delta}}
\,d\xi$, namely, the rate of the lepton energy gain multiplied by the number of the particles 
within the gap. We note that for the parameters used here, the Blandford-Znajek 
reference power is $L_{BZ}=2 \times 10^{43}$ erg\,s$^{-1}$ (see equation~\ref{L_BZ}). 
Hence, the resultant gap luminosity only constitutes a small fraction of the 
Blandford-Znajek jet power. We note that for a very small current value, the gap power can 
deviate significantly from the scaling law derived for thin ($h\ll r_g$) gaps \citep{kats18}.

We were not able to find any steady solution for $\dot{m}>10^{-4.5}$, and 
thus considered here the case where the black hole is fed by accretion at 
a rate $\dot{m}=10^{-5.0}$. We expect higher accretion rates to lead to
gap breakdown due to increased seed injection and efficient pair cascade
development \citep{lev11}. Hence, one can say that steady gaps are not 
allowed at sufficiently high accretion rates. 

If the findings presented in table~\ref{tab_1} are viewed in the context of 
recent VHE observations \citep[see e.g.][for a review]{rie18}, a gap VHE
luminosity of $L_{TeV}/L_{BZ}\sim 5\times 10^{-3}$, as e.g. required for the 
flaring events in M87, would then be indicative of global current values  
$|J_{o}^{*}| \gppr 0.3$. This would suggest that a steady gap model could 
be applied to the VHE activity in M87, providing also a plausible current 
value. As already mentioned, the global magnetospheric current is a critical 
function associated with jet formation.

\begin{figure*}
\gridline{\fig{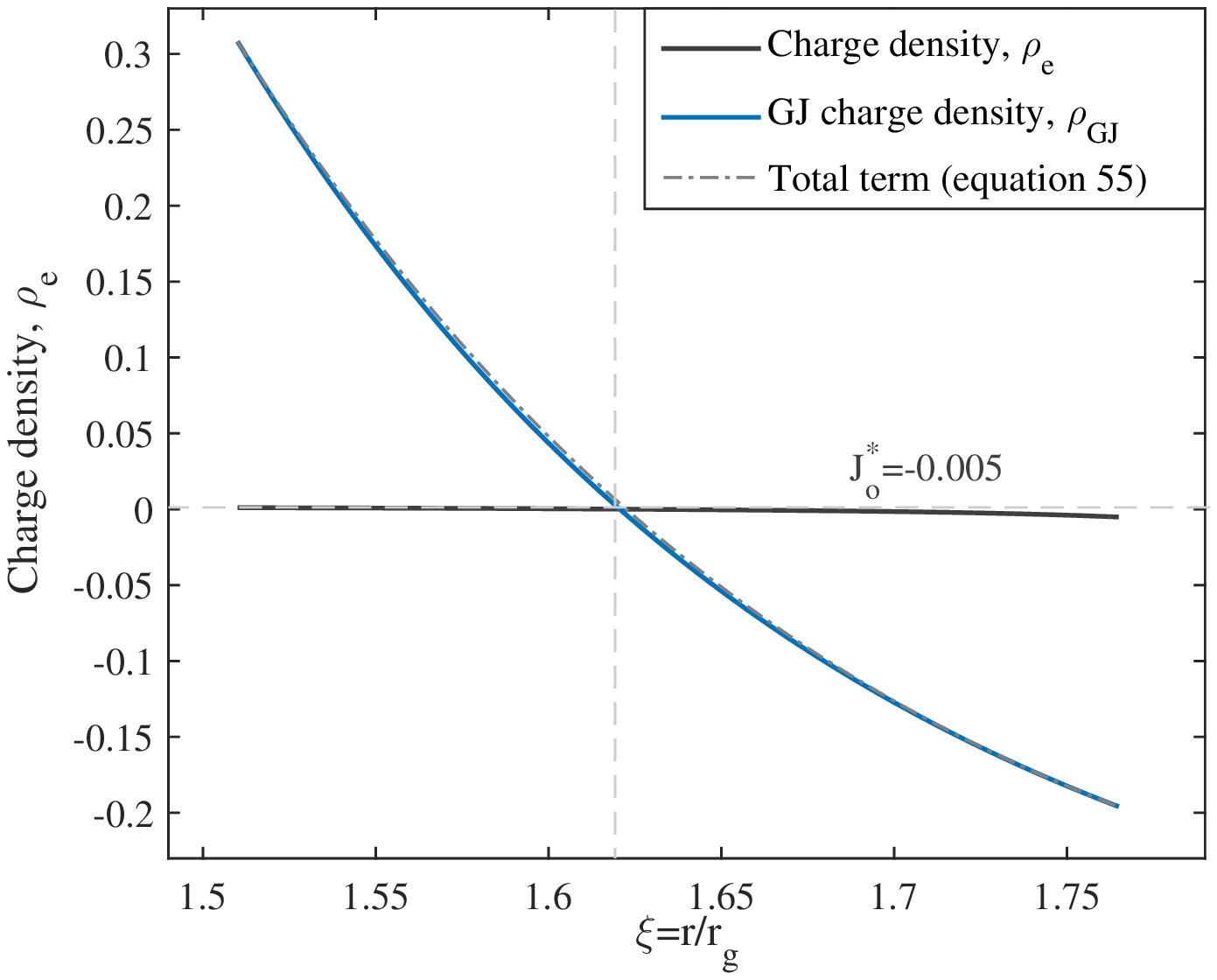}{0.40\textwidth}{(a)}
              \fig{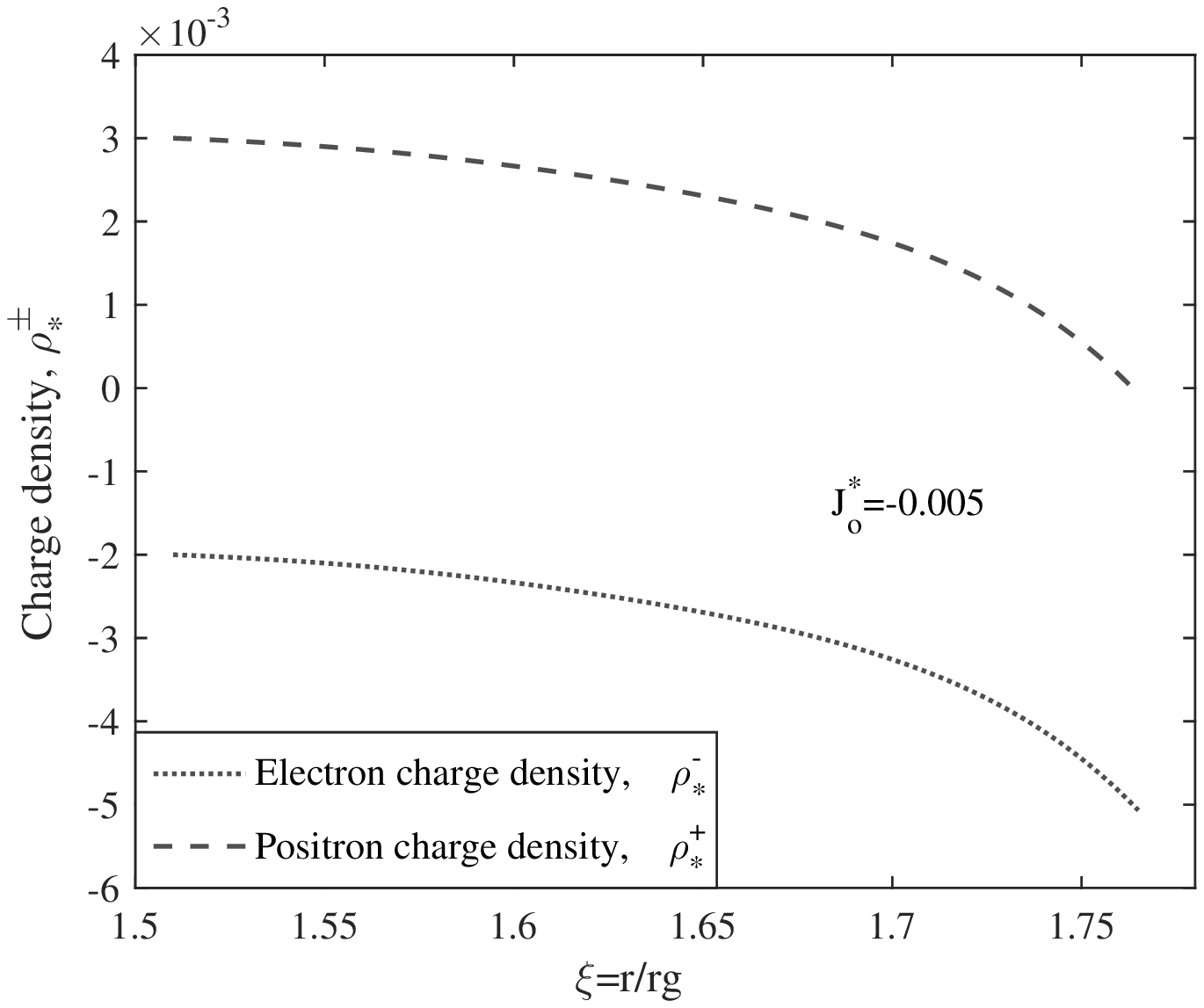}{0.40\textwidth}{(b)}
             }
\gridline{\fig{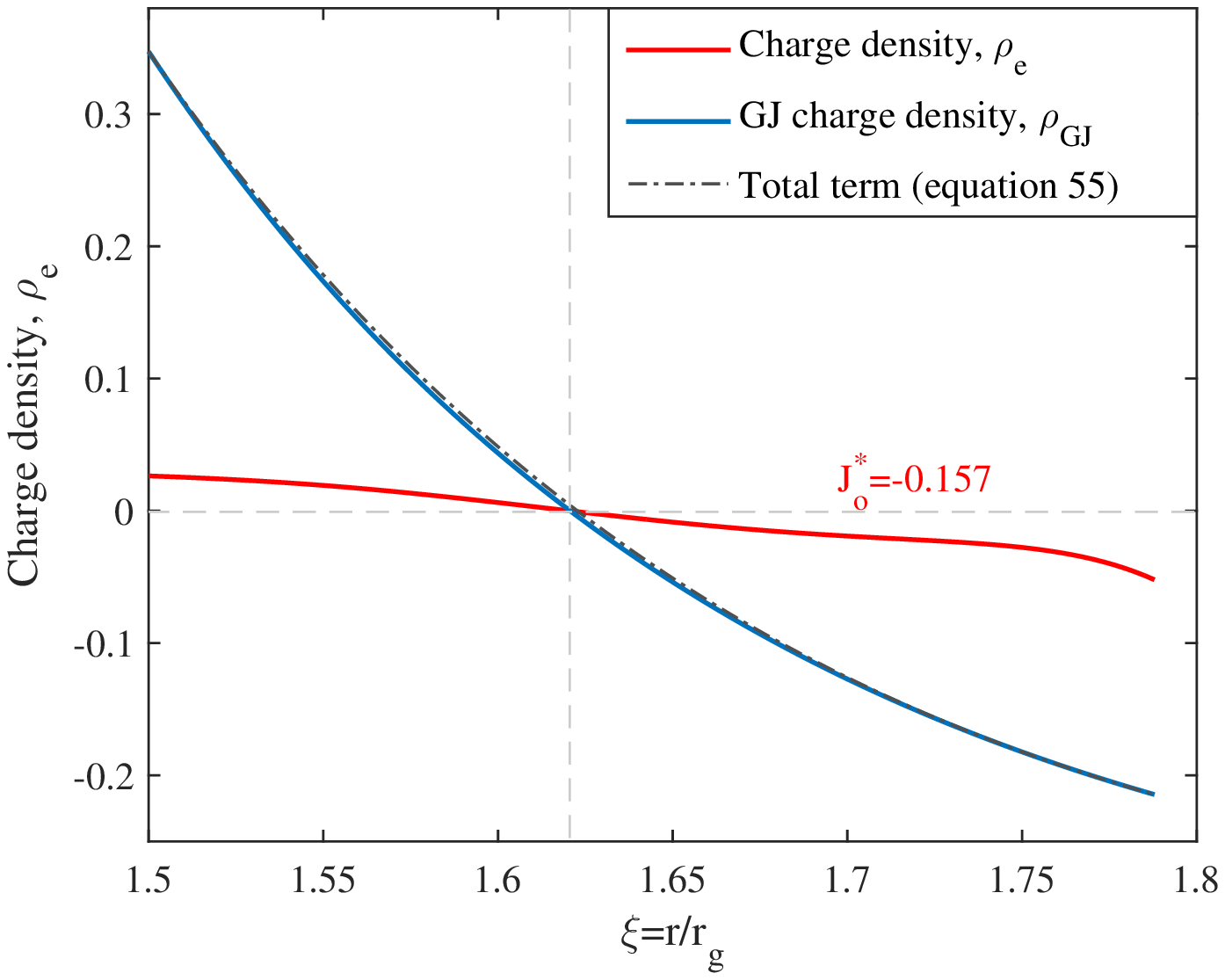}{0.40\textwidth}{(c)}
              \fig{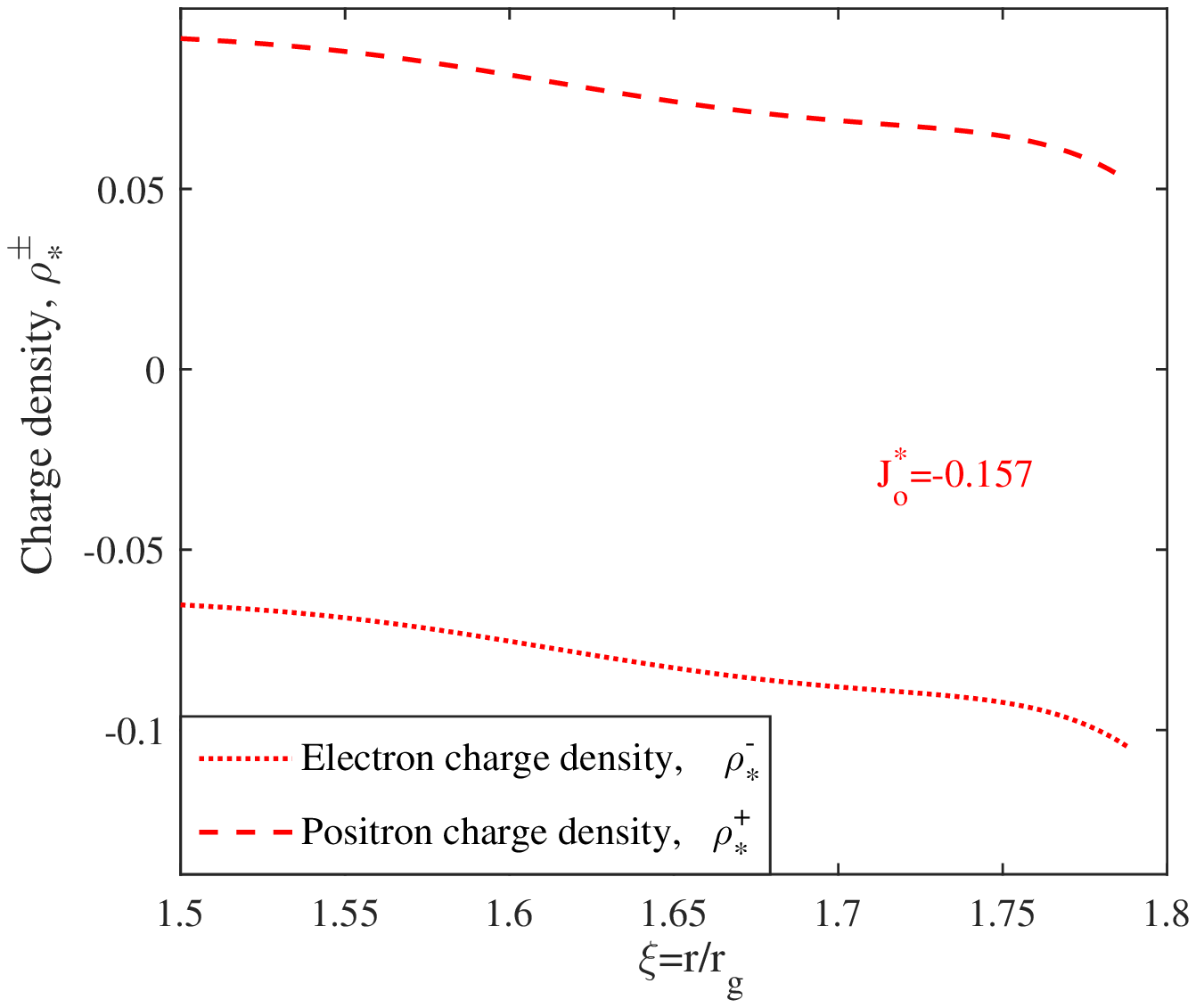}{0.40\textwidth}{(d)}
            }
\gridline{\fig{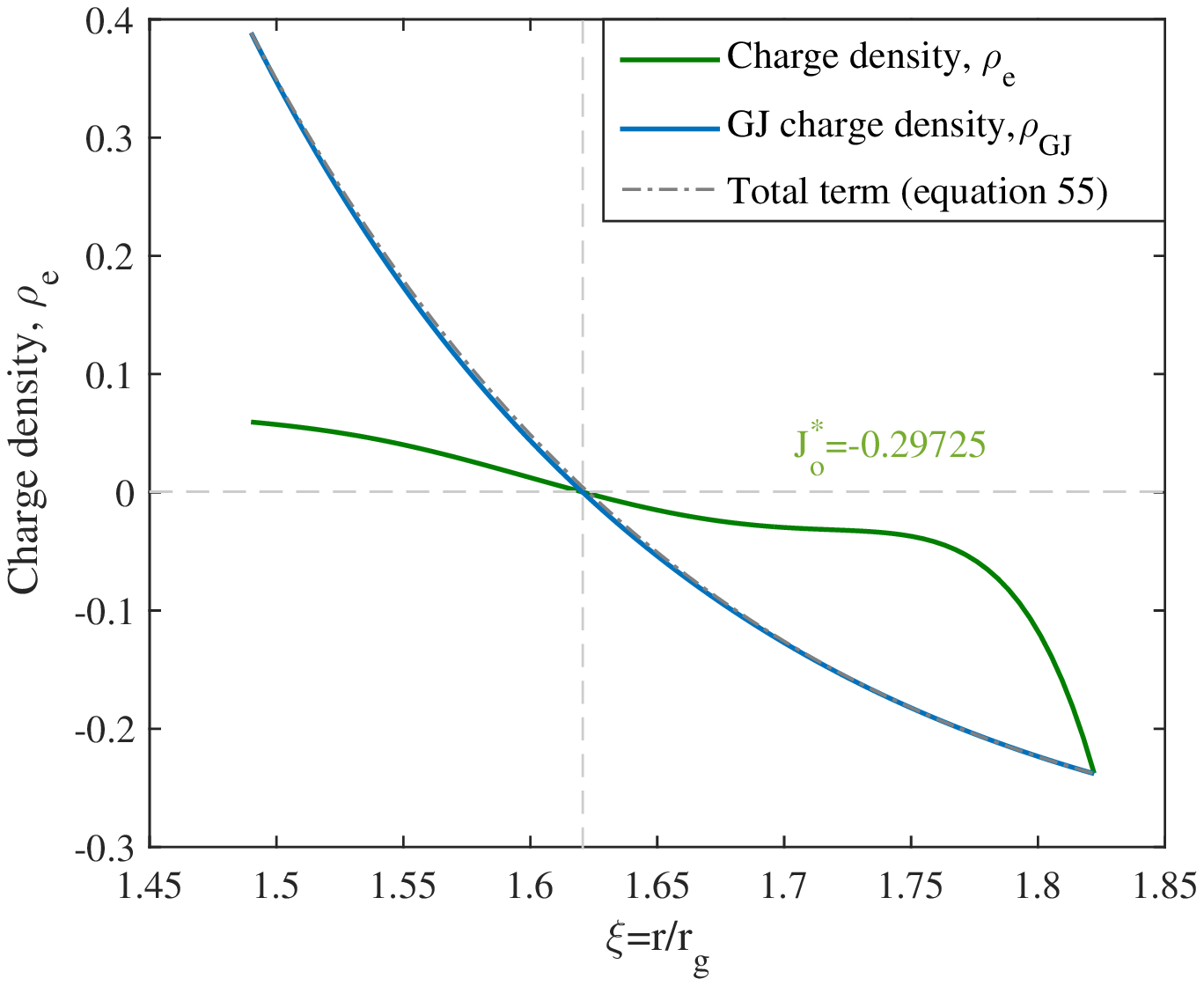}{0.40\textwidth}{(e)}
              \fig{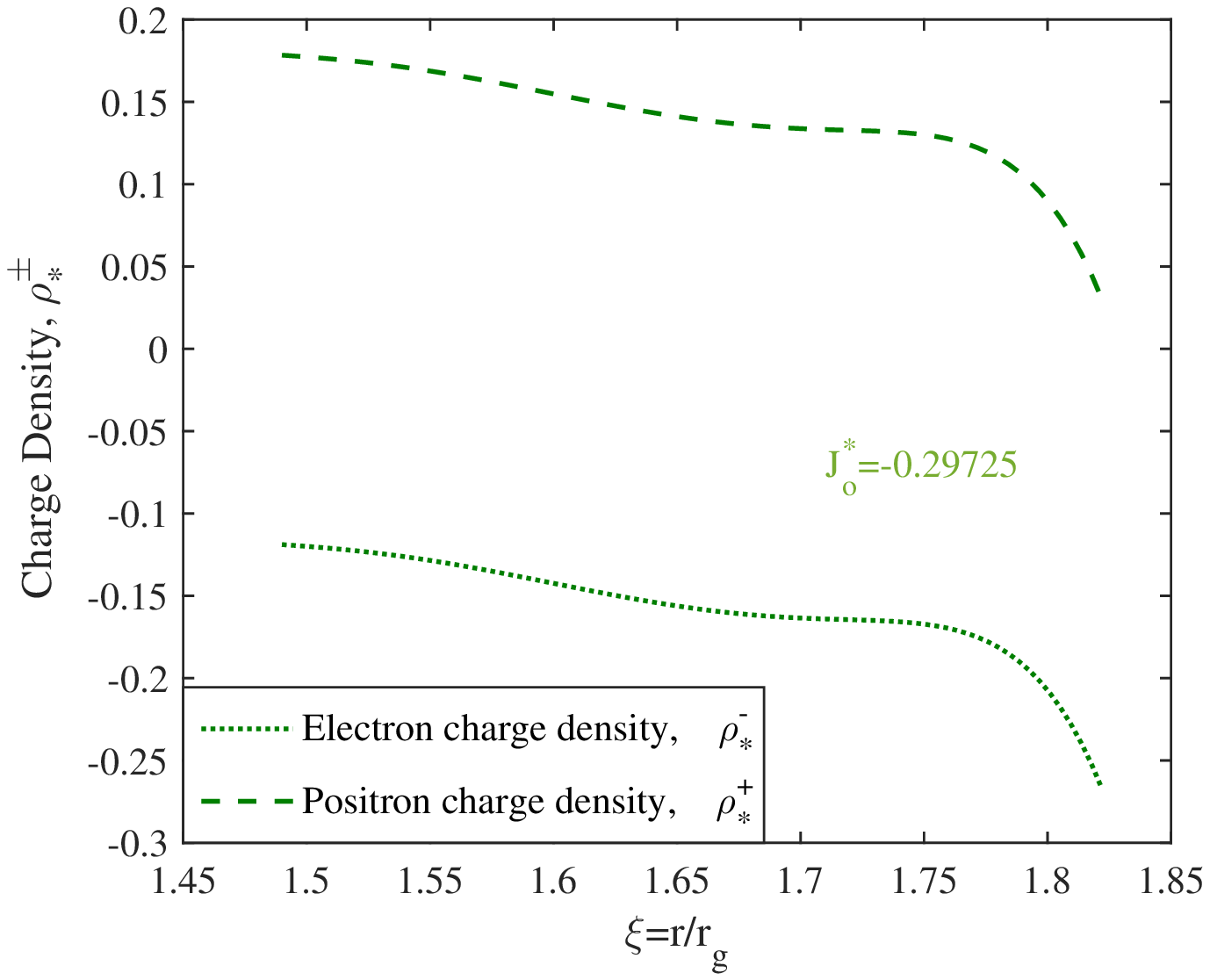}{0.40\textwidth}{(f)}
            }
\vspace*{-0.3cm}
\caption{Illustration of the resulting charge densities with respect to the 
Goldreich-Julian one (left column) and the positron/electron charge densities 
as distributed within the gap region (right column) shown for current value 
$J_{o}^{*}=-0.005$ (black line), $J_{o}^{*}=-0.157$ (red line) and $J_{o}^{*}
=-0.297$ (green line), respectively. The parameters used are: $M_{9}=1.0$, 
$\alpha_{s}^{*}=1.0$, $\dot{m}=10^{-5.0}$ and $\theta=30^{o}$. \label{fig:8.2}}
\end{figure*}

Examples of the total charge density with respect to the Goldreich-Julian one 
and the distribution of positrons and electrons within the gap are presented 
in the left and right columns of figure (\ref{fig:8.2}), respectively. Each row 
in this figure corresponds to different values of the global magnetospheric 
current. The resultant charge density (amount) remains always lower than the 
critical Goldreich-Julian one (left column). Furthermore, the relaxation of 
conditions (\ref{eq5.03}) and (\ref{eq5.04}) has led to gap solutions with 
injection of both species at the boundaries (with the exception of the outer 
boundary for the current $J_{o}^{*}=-0.005$; see right column in figure 
\ref{fig:8.2}). We note that small current values (e.g., $J_{o}^{*}=-0.005$) 
lead to a highly underdense gaps, while higher current values (e.g., $J_{o}^{*}
=-0.29725$) provide a charge distribution that can match the Goldreich-Julian 
charge density at the outer boundary. 

One can use the results presented in figure (\ref{fig:8.2}, right) to verify 
that the global current remains constant along the field line (see equations 
\ref{eq4.08} and \ref{eq4.09}). The distribution of $\gamma$-ray photons, on 
the other hand, can exhibit a complex behaviour. Equation (\ref{eq4.14})
shows that the source term is formed by the summation of outcoming and incoming 
photons in each energy bin. In our case there are some bins which contribute 
decisively to the gap structure and many others which do not. In the numerical
procedure some of the latter may take on negative values, which may indicate 
a generic (possibly structural) problem of a steady gap model. However, since 
their total contribution is negligible, this does not affect the overall results.

\begin{figure}[H]
\includegraphics[width=0.48\textwidth,height=7.4cm]{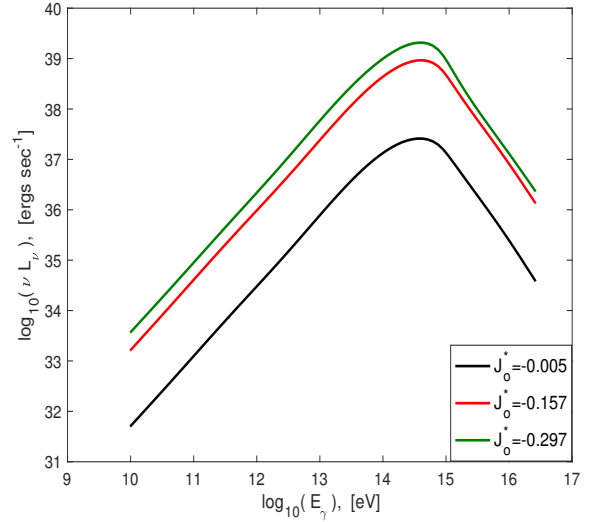}
\caption{Spectral energy distribution of the outgoing gamma-ray photons 
from the gap for different current densities.}\label{fig:SED}
\vspace*{0.5cm}
\end{figure}

Figure~(\ref{fig:SED}) shows the IC-dominated spectral energy distribution 
(SED) of the outgoing photons at the end of the gap, i.e., $\nu L_{\nu} = 4\,\pi\,\,r_2^2\, 
c\, E_{\gamma}^2\, P_{\gamma}^+(r_2,E_{\gamma})$. At lower energies ($\sim 1$ GeV),
curvature emission, which is not shown here, will dominate the spectrum. Note that
this gap spectrum will be reprocessed by absorption, with the resultant spectrum 
further modified by secondary pair emission outside the gap \citep[e.g.,][]{hirpu16}.

\subsection{Solutions for fixed global current} \label{subsec:08.02}

In the previous subsection, gap solutions for a fixed accretion rate and different 
choices of the global current were explored. Here, we keep the current constant, 
seeking to investigate structural variations of the gap due to changes in the 
accretion rate. Three different values of the accretion rate are explored, namely, 
$\dot{m}=10^{-5.0}$, $\dot{m}=10^{-6.0}$ and $\dot{m}=10^{-6.5}$.

\begin{figure*}[htb]
\gridline{\fig{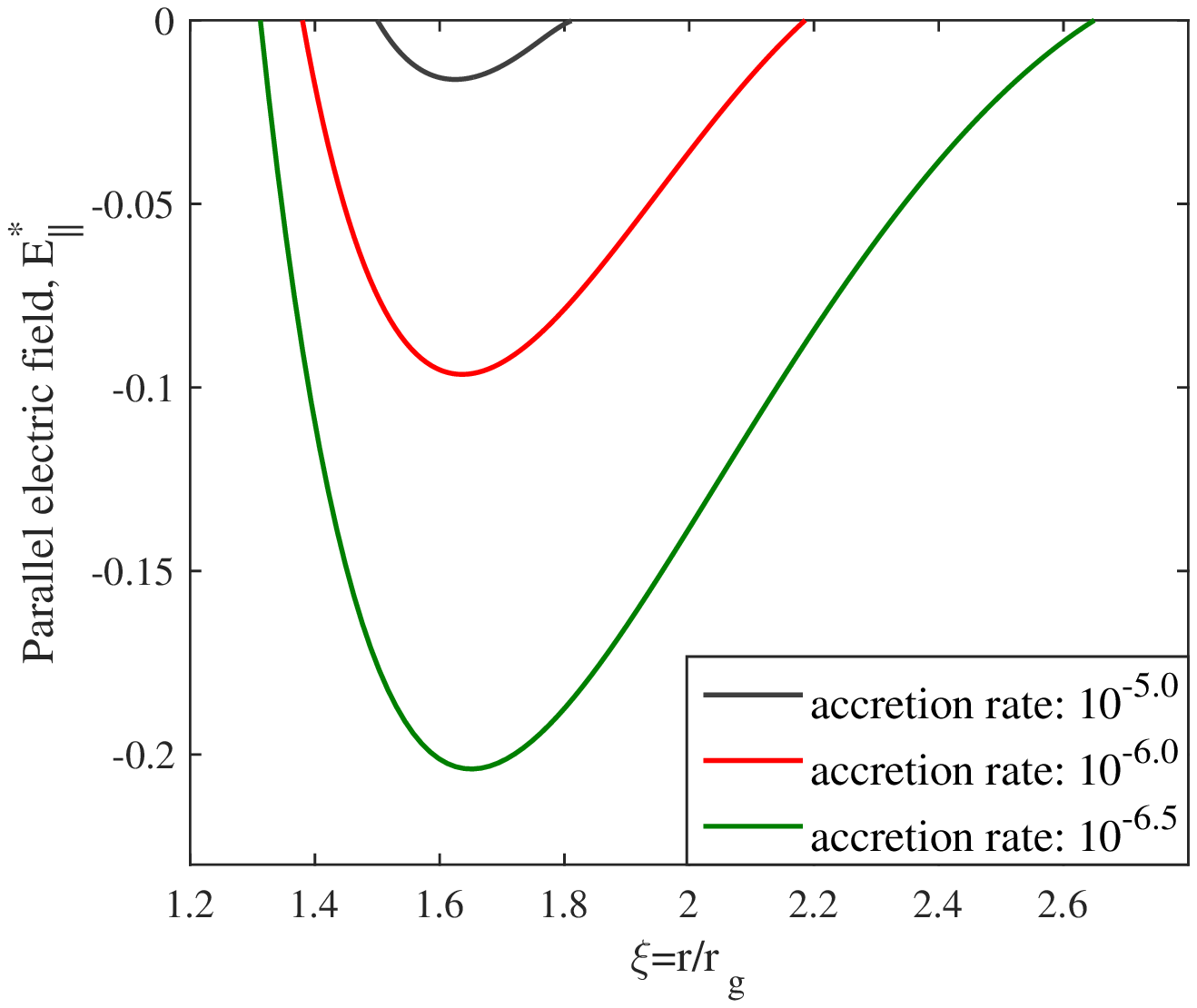}{0.4\textwidth}{(a)}
              \fig{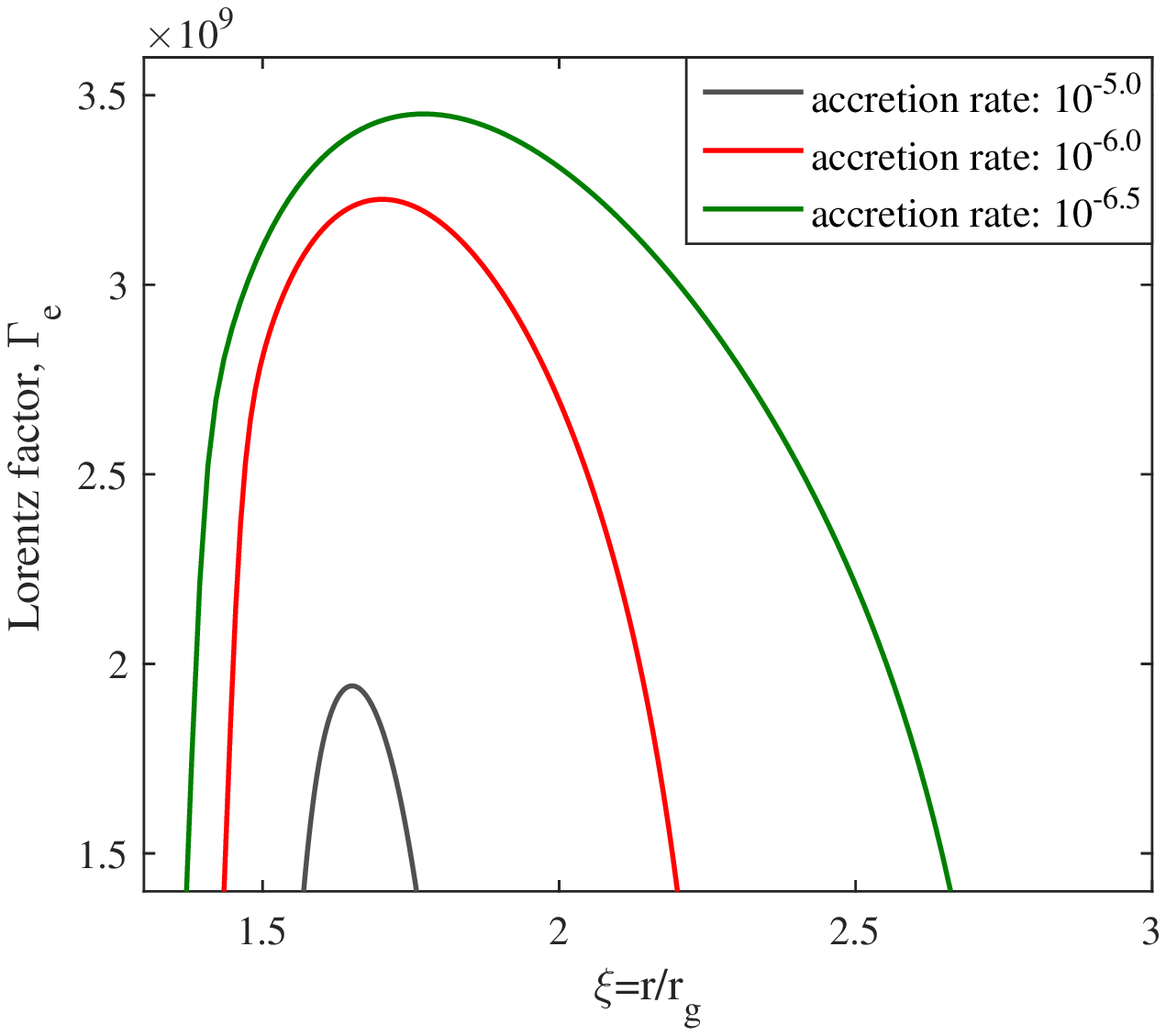}{0.4\textwidth}{(b)}
             }
\gridline{\fig{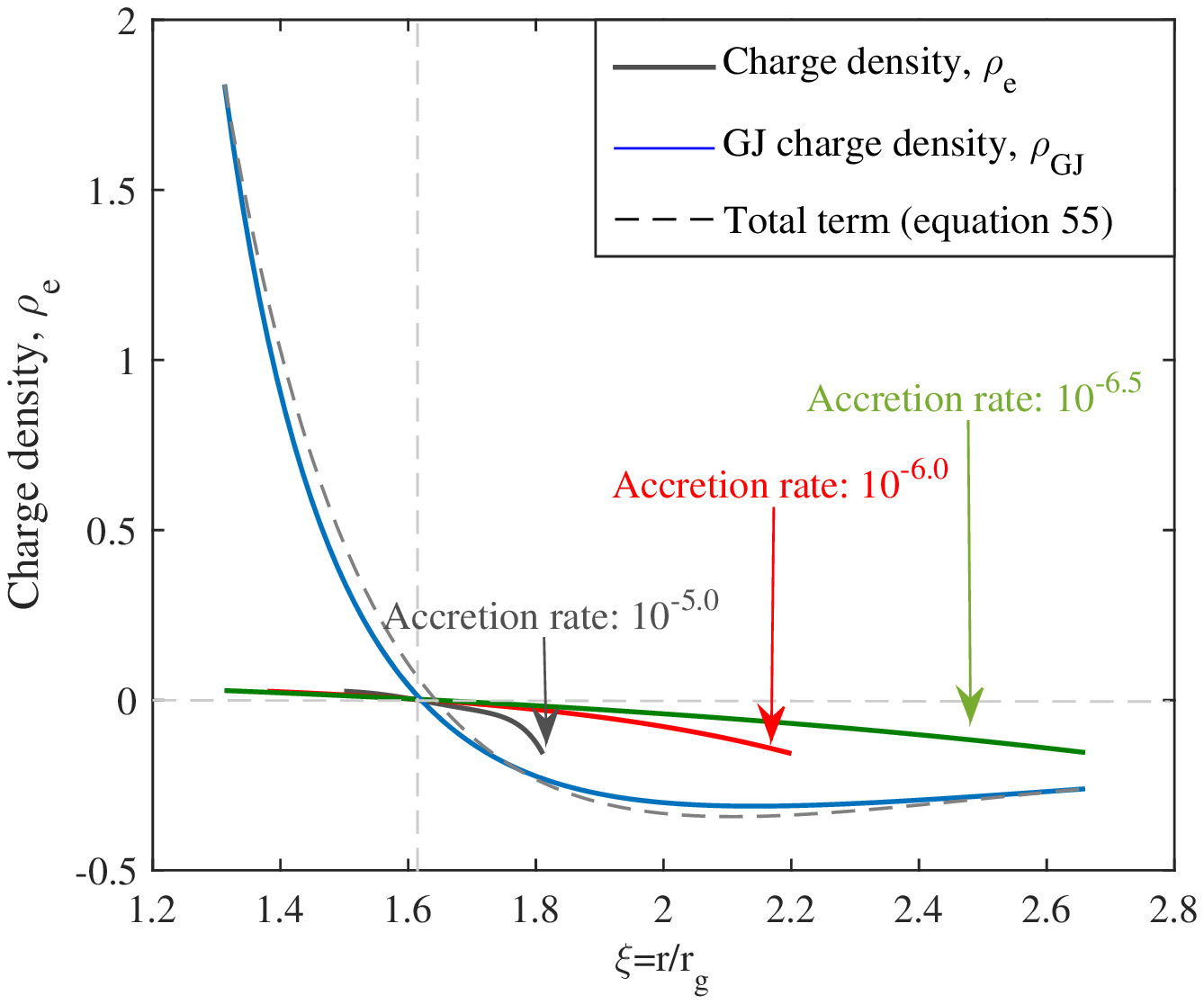}{0.4\textwidth}{(c)}
              \fig{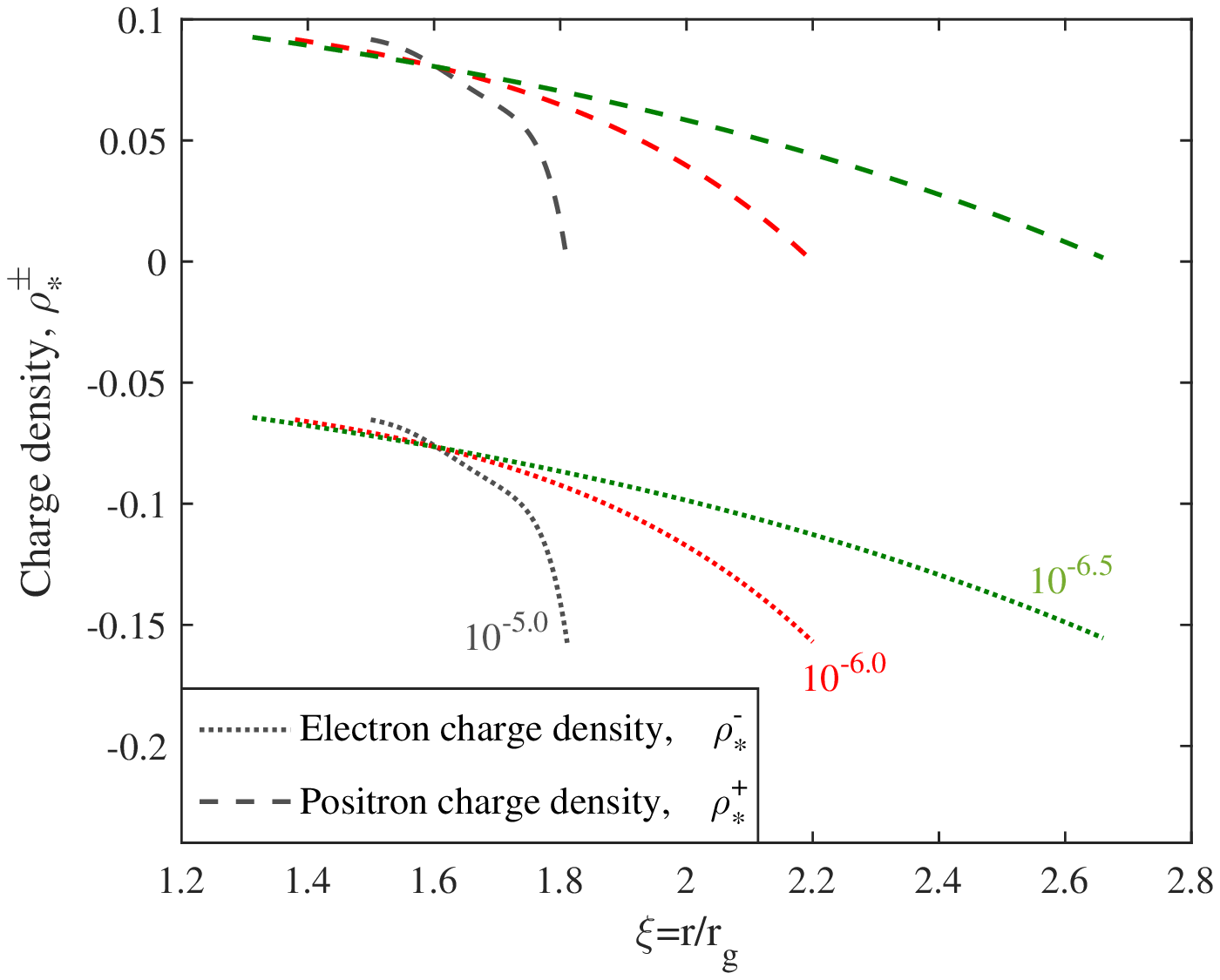}{0.4\textwidth}{(d)}
            }
\caption{Distribution of the parallel electric field component $\mathcal{E}_{||}^{*r}$ 
(top left), the particle Lorentz factor $\Gamma_{e}$ (top right), 
the total charge density $\rho_{e}$ along with the Goldreich-Julian one (bottom left), 
and the positron/electron charge densities $\rho_{*}^{\pm}$ (bottom right) for a 
fixed global current of $J_{o}^{*}=-0.157$ and accretion rates $\dot{m}=10^{-5.0}$ 
(black line), $\dot{m}=10^{-6.0}$ (red line), and $\dot{m}=
10^{-6.5}$ (green line), respectively.
\label{fig:8.4}}
\end{figure*}

\begin{deluxetable*}{ccccc}[b!] 
\tablenum{2}
\tablecaption{Gap properties for fixed global current}\label{tab_2}
\tablewidth{0pt}
\tablehead{
\colhead{Accretion Rate} & \colhead{Gap Size} & \colhead{Voltage Drop} 
& \colhead{BZ power} & \colhead{Gap power}  \\
\colhead{$\dot{m}=\dot{M}/\dot{M}_{Edd}$} & \colhead{$h/r_{g}$} 
& \colhead{$\times 10^{16}$ V} & \colhead{$\times 10^{43}\,erg\,s^{-1}$} 
& \colhead{$\times 10^{40}\,erg\,s^{-1}$}  
}
\decimalcolnumbers
\startdata
$10^{-5.0}$ & $0.2879$ & $6.3$ & $2.0$ & $2.2$ \\ 
$10^{-6.0}$ & $0.8200$ & $29.1$ & $0.2$ & $5.5$ \\ 
$10^{-6.5}$ & $1.3475$ & $50.1$ & $0.06$ & $5.9$ \\ 
\enddata
\tablecomments{Results for the gap extension, the associated voltage drop and 
total gap power for a fixed global current of $J_{o}^{*}=-0.157$, along with 
the Blandford-Znajek reference power (equation~\ref{L_BZ}).}
\end{deluxetable*}

Figure (\ref{fig:8.4}) presents examples for the distribution of the parallel electric 
field component (top left diagram), the Lorentz factor of the accelerated pairs 
(top right diagram), the total charge density along with the Goldreich-Julian one 
(bottom left diagram) as well as the positron and electron charge densities within 
the gap (bottom right diagram). The current value has been fixed to $J_{o}^{*}=
-0.157$ and the BH parameters are $M_{9}=1.0$, $\alpha_{s}^{*}=1.0$, and 
$\theta=30^{o}$.

As can be seen in Figure (\ref{fig:8.4}), the gap extension increases as the accretion 
accretion rate decreases (see the top left panel). This is related to the fact that 
for lower soft photon fields the pair production efficiency is reduced such that larger 
gaps are expected. The gap size is roughly comparable to the gravitational radius for 
the lower accretion rates considered (see table \ref{tab_2} for details). Maximum 
Lorentz factors ($\Gamma_{e}\sim 2.0-3.5\times 10^9$) are achieved beyond the 
extremum of the electric field (see the top right panel). Evidently, the lower the 
accretion rate, the higher the particle Lorentz factor. The resultant charge density 
satisfy $|\rho_{e}|\leq |\rho_{GJ}|$ everywhere (see the 
bottom left panel). Finally, figure (\ref{fig:8.4}) shows that charge injection of both 
species (i.e, relaxation of condition \ref{eq5.03} and \ref{eq5.04}) has taken place 
mostly at the inner boundary (see the bottom right panel).

The attainable gap luminosities are calculated in table (\ref{tab_2}). Accordingly, 
only a fraction of the Blandford-Znajek power is released by the gap accelerator. 
In the case of M87, for example, an accretion rate of $\sim 10^{-5.0}$ seems to be 
required. It is worth noting that this value is compatible with recent observational 
estimates for M87 \citep{aki19b}.

\begin{figure}[htb]
\includegraphics[width=0.48\textwidth]{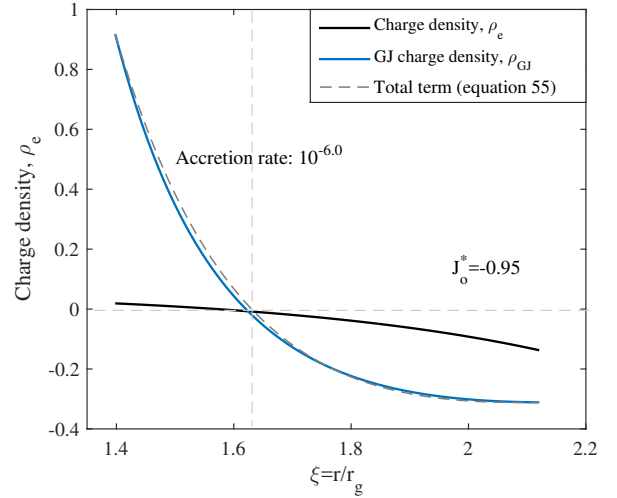}
\caption{Charge distribution for a global magnetospheric current $J_{o}^{*}
=-0.95$ and accretion rate $\dot{m}=10^{-6.0}$.}
    \label{figure_4Gd}
\end{figure}

As argued above, the existence of steady gap solutions is possible even for high 
values of the global magnetospheric current if charge injection of both species 
is allowed to occur at the gap boundaries (i.e., relaxation of conditions 
\ref{eq5.03} and \ref{eq5.04}). Table~(\ref{tab_3}) provides one example with 
$J_o^* \sim 1$, assuming $\dot{m}=10^{-6.0}$ for which $L_{BZ}=2\times10^{42}$ 
erg s$^{-1}$ (equation~\ref{L_BZ}). The resultant charge distribution is shown in 
figure~\ref{figure_4Gd}. Finally, we note that for a high current value no steady gap 
solution could be determined for accretion rates much higher than $\sim 10^{-6.0}$.

\begin{deluxetable}{cccc}[H]
\tablenum{3}
\tablecaption{Gap properties for higher current value} \label{tab_3}
\tablewidth{0pt}
\tablehead{
\colhead{Global Current} & \colhead{Gap Size} & \colhead{Voltage Drop} 
& \colhead{Gap Power}  \\
\colhead{$J_{o}^{*}=J_{o}/c\,\rho_{c}$} & \colhead{$h/r_{g}$} 
& \colhead{$\times 10^{16}$ V} & \colhead{$\times 10^{40}\,erg\,s^{-1}$} 
}
\decimalcolnumbers
\startdata
$-0.95$ & $0.7225$ & $21.9$ & $4.4$ \\
\enddata
\tablecomments{Results for the gap extension, the associated voltage drop and 
total gap power for a global current $J_0^*\sim 1$, assuming $\dot{m}=10^{-6.0}$.}
\end{deluxetable}

\section{Discussion}
\begin{figure}[h!]
\includegraphics[width=0.49\textwidth,height=7.0cm]{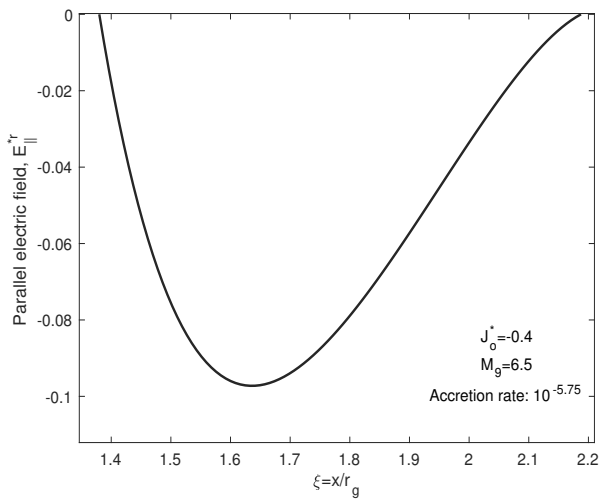}
\caption{Exemplary parallel electric field distribution for the case of 
M87.}\label{figure_4K}
\end{figure}

The above calculations support the notion that pair cascades in magnetospheric 
gaps can ensure field screening and lead to a detectable gamma-ray contribution 
in nearby, underluminous and misaligned AGNs \citep[e.g.,][]{rie19}. The radio galaxy 
M87 represents a prime candidate in this regard. Its variable (day-scale) VHE
activity could possibly be related to magnetospheric processes 
and provide a signature of jet formation \citep{lev11,kats18}. In order to 
explore this in more detail, we performed exemplary model calculations using the 
latest BH mass estimate of $M_9=6.5$ \citep{aki19a,aki19c}. The results are shown 
in Table~\ref{tab_4} and Fig.~\ref{figure_4K}. In this case, gap sizes of the 
order of $\sim 0.8\,r_g$ are obtained, suggesting that its VHE emission could be 
variable down to timescales of $\sim0.4$ days. The inferred gap power of $\sim 5 
\times 10^{41}$ $erg\,s^{-1}$ would make it in principle possible to accommodate 
the VHE emission seen during its high states \citep[e.g.,][]{aliu12,ait19}. These 
results provide tentative support for a gap origin of the VHE emission in M87, though 
detailed spectral modeling will be needed in the end. The
accretion rate employed for this calculation is close to the mean MAD value used 
in GRMHD simulations \citep{aki19b}, and would correspond to jet powers of a few 
times $10^{43}$ $erg\,s^{-1}$. We note that since in our model the gap width is 
primarily determined by the accretion rate, observations of rapid VHE variability 
could in principle be used to impose an lower limit on the 
accretion rate. The estimated voltage drop for M87 is of the order $\sim10^{18}$ V
(Table~\ref{tab_4}), suggesting that proton acceleration is limited to $\sim10^{18}$ 
eV. Hence, if gap-type particle acceleration is associated with ultra-high-energy 
cosmic-ray (CR) production, the CR composition might be expected to become 
heavier toward highest energies. This seems compatible with current Pierre Auger 
results \citep[e.g.,][]{bat2019}, though we note that the conditions in M87 are rather 
exceptional, making a generalization somewhat difficult.

\begin{deluxetable}{cccc}[h!]
\tablenum{4}
\tablecaption{Gap properties as inferred for M87} \label{tab_4}
\tablewidth{0pt}
\tablehead{
\colhead{Global Current} & \colhead{Gap Size} & \colhead{Voltage Drop} 
& \colhead{Gap Power}  \\
\colhead{$J_{o}^{*}=J_{o}/c\,\rho_{c}$} & \colhead{$h/r_{g}$} 
& \colhead{$\times 10^{17}$ Volts} & \colhead{$\times 10^{41}\,erg\,s^{-1}$} 
}
\decimalcolnumbers
\startdata
$-0.4$ & $0.8076$ & $9.8$ & $4.9$ \\
\enddata
\tablecomments{Results for the gap extension, the associated voltage drop and 
total gap power for a global current $J_0^*=-0.4$, assuming $M_{9}=6.5$, and 
$\dot{m}=10^{-5.75}$.}
\end{deluxetable}

A straightforward comparison of our study with previous works is complicated by 
the fact that steady 1D gaps have been studied for different regimes (e.g., 
related to choices of the physical frame, the soft photon field, and the charge 
and photon boundary conditions). Our results, nevertheless, verify earlier 
findings. The gap widths, the particle Lorentz factors and the voltage differences 
obtained here agree with recent BH studies \citep[e.g.,][]{hirpu16, hir16,lev17}. 
This can be understood taking into account that steady gaps are eventually mainly 
regulated by the disk accretion rate. In accordance with \citet{lev17} and 
\citet{hir16}, we also find that the gap extension becomes larger with increasing 
the global magnetospheric current (e.g., see Fig. \ref{fig:8.1}), and that the gap 
luminosity increases as the accretion rate is decreased (see Table \ref{tab_2}). 
Differences in the shape of the parallel electric field curves appear attributable 
to slightly different boundary conditions (we recall that we have imposed $|\rho_{e}|
\leq |\rho_{GJ}|$ everywhere in the gap). We are thus confident that the approach 
adopted here leads to reasonable steady gap solutions for plausible current values 
and a useful estimation of the gap extension and associated voltage drop.

While the present 1D model allows us to get insights into the physics
characteristics of magnetospheric gaps in AGNs, its limitations should 
be kept in mind. This includes the usage of a monopole magnetic field
structure, a simplified description of the ADAF soft photon field,
and the application of special relativity in, e.g., the equation of 
motion. The latter, however, does not seem to introduce significant
differences when the findings are compared with more general models
\citep[e.g.,][]{lev17}. As common to steady approaches, the underlying
framework treats gaps as not affecting the global magnetospheric
structure, which may formally only be valid for thin gaps. To improve 
upon these limitations requires suitable extension and time-dependent
modeling \citep[e.g.,][]{lev05}, which we plan to address in a future 
work.

The extent to which gap formation may be intermittent is not clear. While
instructive, recent PIC simulations are not yet conclusive in this regard.
In the 1D simulations by \citet{che18} for example, gaps are dynamically 
formed as pairs are advected out of the system \citep[cf. also][]{che19}. 
The solutions are highly time dependent with no steady gaps being seen, 
and reveal quasi-periodic gap opening on timescale $\sim r_g/c$. Their 
results suggest that gaps can develop "everywhere" and extend over several 
$10\%$ of $r_g$ or more. Their simulations, however, employ a rather high 
minimum energy for the soft photon distribution ($\epsilon_{\rm s,min} =0.5$ 
eV) with the possible caveat that the Klein-Nishina regime for IC scattering 
is quickly reached, and the pair creation length becomes large compared with 
the Thomson mean free path. This might partly explain why unsteady, extended 
gaps are observed. In the GR simulations by \citet{lev18}, on the other hand, 
an approach to a quasi-steady state, characterized by rapid, small-amplitude 
$E_{\|}$-oscillations and self-sustained pair cascades resulting in
quasi-stationary pair and gamma-ray spectra, is seen. Longer runs may be 
needed to better understand the differences. While both simulations 
employ simplified (fixed single power-law) soft photon descriptions, they
use different low-energy cutoffs $\epsilon_{\rm s,min}$ (viz., $10^{-6}$ 
vs. $10^{-8}\,m_e c^2$) and explore different regimes (e.g., high vs low 
Thomson mean free path), and are thus not straightforward to compare.
Our approach chosen here seems beneficial 
in that it allows us to get first insights into possible dependencies of the 
gap structure on different and more complex ambient soft photon fields, 
including their variation with accretion rate. In particular, both
the low-energy peak of the ADAF emission ($\nu_p^{syn} \propto \dot{m}^{3/4} 
T_e^2$) and the shape of the (Comptonized) emission above it ($F_{\nu,s} 
\propto \nu_s^{-\Gamma}, \Gamma \sim\,2.2-2.6$) depend on the accretion rate 
of the source. If one supposes that gap-accelerated electrons are able to reach, 
e.g., $\Gamma_e \sim 10^9$, then the IC power is dominated by 
upscattering of photons with $\nu_s \sim 10^{10-12}$ Hz. 
Similarly, as VHE photons preferentially interact with soft photons of 
energy $\epsilon_s \sim 0.01~(100~\mathrm{TeV}/\epsilon_{\gamma})$ eV, 
the low-energy part of the soft photon distribution becomes relevant as well. 
This suggests that a suitable choice of $\epsilon_{\rm s,min}$
(and related energy density) is relevant for steady gap formation.
In general, for comparable simulations it seems important to employ a soft 
photon field such that over a wide energy range efficient pair creation is 
ensured within the simulation box.

In principle, gaps represent an essential part of the 
global magnetospheric structure. A self-consistent analysis thus requires 
a proper treatment of the coupling between the gap and the force-free 
region of the outflow. This will eventually require global long-term GR 
plasma simulations \cite[see][for first attempts]{par19,cri20}, incorporating 
radiative processes and back-reaction, as well as realistic astrophysical 
boundary conditions. Given the complexity of the problem (often requiring 
nontrivial rescaling), local gap solutions, in which the magnetospheric 
current is treated as free parameter, can be complementarily sought for 
to explore some of the physics characteristics. 
Such an approach, as also chosen here, implicitly assumes that the global 
magnetospheric structure (i.e., magnetic field geometry and angular velocity 
of magnetic surfaces) is not significantly affected by the gap activity, 
which introduces limitations. In the present paper we have explored current 
values for which steady gap closure (with $|\rho_e|\leq |\rho_{GJ}|$) around 
the null surface in a realistic accretion environment can be achieved.
Intermittent gap activity could possibly facilitate higher charge 
multiplicities (higher global current values), though this seems at the 
same time to be accompanied by a decrease in gap extension (i.e., $h/r_g 
\ll 1$ in \citealt{lev18}). Whether quasi-steady gaps can exist in a global 
setup (with an inner null and an outer stagnation surface) remains unclear 
\citep[e.g.,][]{lev17}. It seems possible that in a global framework the 
gap activity becomes highly time dependent, possibly revealing some cyclic or 
fast oscillatory behavior in which the gap width (electric field amplitude) 
might be regulated by pair creation balancing pair escape \citep{lev18}. 
This could result in a reduced power output compared to the steady case. A 
straightforward comparison is, however, complicated, due to the use of 
different setups (e.g., no or some charge injection from outside) and 
input parameters (e.g., soft photon description). 

At the conceptual level, efficient pair creation in magnetospheric gaps 
can provide a physical mechanism to guarantee the plasma source and currents 
needed to electromagnetically extract the rotational energy of the black hole. 
The resultant gamma-ray emission is of interest by allowing a unique probe 
of the near-black hole environment.

\section{Conclusion} \label{discussion}
In the present work, a detailed analysis of steady gap acceleration across the null 
surface of a rotating BH magnetosphere embedded in an ADAF soft photon field 
has been presented. The system of equations governing the gap accelerator (e.g. the
radial distributions of the parallel electric field $\mathcal{E}_{||}^{r}$ and the 
charge densities $\rho_e^{\pm}$) has been numerically solved by means of a shooting 
method. 
Gap solutions, assuming suitable boundary conditions (e.g., $\mathcal{E}_{||}^{r}=0$),
are presented for different choices of the global current and BH accretion rate. The 
model has been adjusted to explore the parameter space relevant for low-luminosity 
AGNs such as radio galaxies. The existence of steady gap solutions for high values 
of the global current is shown to be possible if charge injection is allowed at the gap 
boundaries. The extent to which BH gap activity rather follows a highly intermittent 
behavior requires global radiative plasma simulations with realistic input parameters
and boundary conditions.
Our current findings provide support to the notion that the variable VHE emission 
in M87 could arise in the immediate vicinity of its central BH. Future VHE 
observations may thus allow it to probe deeper into the physics of supermassive 
BHs.

\acknowledgments
We appreciate constructive comments by the referee.
We are grateful to Amir Levinson, Alexander Chen, Christian Fendt, Kouichi Hirotani, 
John Kirk, and Felix Aharonian for helpful discussions. We thank Amir Levinson for 
comments on an earlier version of this paper. F.M.R. acknowledges funding by a DFG 
Heisenberg Fellowship under RI 1187/6-1.

\bibliography{paper}{}
\bibliographystyle{aasjournal}

\end{document}